\documentclass[twocolumn]{aastex701}
\usepackage[utf8]{inputenc}
\usepackage{hyperref}
\usepackage{tabularx}
\usepackage{multirow}
\usepackage{amsmath}
\usepackage{amssymb} 

\newcolumntype{Y}{>{\centering\arraybackslash}X}

\received{--}
\revised{--}
\accepted{--}
\submitjournal{AAS}
\shorttitle{SN~2017gkk}
\shortauthors{B. Kumar et al.}

\begin{document}
\title{SN~2017gkk: A moderately luminous and fast evolving Type IIb supernova}
\correspondingauthor{Brajesh Kumar, Jiewei Zhao, Xueling Du} 
\email[show]{brajesh@ynu.edu.cn, meow.jiewei.zhao@gmail.com,\\xldu@stu.ynu.edu.cn}
\author[0000-0001-7225-2475]{Brajesh Kumar}
\affiliation{South-Western Institute for Astronomy Research, Yunnan University, Kunming, Yunnan 650500, People's Republic of China}
\affiliation{Yunnan Key Laboratory of Survey Science, Yunnan University, Kunming, Yunnan 650500, People's Republic of China}
\affiliation{Indian Institute of Astrophysics, II Block, Koramangala, Bengaluru-560034, Karnataka, India}
\email{brajesh@ynu.edu.cn, brajesharies@gmail.com}
\author[0000-0002-6688-0800]{D. K. Sahu}
\affiliation{Indian Institute of Astrophysics, II Block, Koramangala, Bengaluru-560034, Karnataka, India}
\email{dks@iiap.res.in}
\author[0000-0003-2091-622X]{Avinash Singh}
\affiliation{The Oskar Klein Centre, Department of Astronomy, Stockholm University, AlbaNova, SE-10691 Stockholm, Sweden}
\email{avinash21292@gmail.com}
\author[0009-0009-8633-8582]{Jiewei Zhao}
\affiliation{South-Western Institute for Astronomy Research, Yunnan University, Kunming, Yunnan 650500, People's Republic of China}
\affiliation{Yunnan Key Laboratory of Survey Science, Yunnan University, Kunming, Yunnan 650500, People's Republic of China}
\email{meow.jiewei.zhao@gmail.com}
\author[0000-0003-3533-7183]{G. C. Anupama}
\affiliation{Indian Institute of Astrophysics, II Block, Koramangala, Bengaluru-560034, Karnataka, India}
\email{gca@iiap.res.in}
\author[0009-0005-5865-0633]{Xueling Du}
\affiliation{South-Western Institute for Astronomy Research, Yunnan University, Kunming, Yunnan 650500, People's Republic of China}
\affiliation{Yunnan Key Laboratory of Survey Science, Yunnan University, Kunming, Yunnan 650500, People's Republic of China}
\email{xldu@stu.ynu.edu.cn}
\author[0000-0001-6706-2749]{Mridweeka Singh}
\affiliation{Indian Institute of Astrophysics, II Block, Koramangala, Bengaluru-560034, Karnataka, India}
\email{yashasvi04@gmail.com}
\author[0009-0007-9727-7792]{Hrishav Das}
\affiliation{Indian Institute of Astrophysics, II Block, Koramangala, Bengaluru-560034, Karnataka, India}
\affiliation{Pondicherry University, Chinna Kalapet, Kalapet, Puducherry 605014, India}
\email{hrishav.das@iiap.res.in}
\author[0009-0008-9062-1455]{Amrit Dutta}
\affiliation{Indian Institute of Astrophysics, II Block, Koramangala, Bengaluru-560034, Karnataka, India}
\affiliation{Pondicherry University, Chinna Kalapet, Kalapet, Puducherry 605014, India}
\email{amrit.dutta@iiap.res.in}
\author[0009-0009-4872-1134]{Gitika Rameshan}
\affiliation{Indian Institute of Astrophysics, II Block, Koramangala, Bengaluru-560034, Karnataka, India}
\affiliation{Academy of Scientific and Innovative Research (AcSIR), Ghaziabad, Uttar Pradesh, 201002, India}
\email{gitika.rameshan@iiap.res.in}
\author[0000-0002-0525-0872]{Rishabh Singh Teja}
\affiliation{Tsung-Dao Lee Institute, Shanghai Jiao Tong University, No.1 Lisuo Road, Pudong New Area, Shanghai, People's Republic of China}
\email{rsteja001@gmail.com}
\author[0009-0008-6974-8611]{Yang Yang}
\affiliation{South-Western Institute for Astronomy Research, Yunnan University, Kunming, Yunnan 650500, People's Republic of China}
\affiliation{Yunnan Key Laboratory of Survey Science, Yunnan University, Kunming, Yunnan 650500, People's Republic of China}
\email{yyang@stu.ynu.edu.cn}

\begin{abstract}
We present optical imaging (\textit{BVRI} bands) and low-resolution optical spectroscopic follow-up of Type IIb supernova SN~2017gkk that occurred in NGC~2748. These observations consist of 31 and 17 epochs of photometry and spectroscopy, spanning $\sim$\,218 days since the $V$-band peak. The light curve analysis infers that the SN reached maximum in $\sim$\,13~days after the explosion with an absolute magnitude of $M_\mathrm{V}$\,=\,--17.01\,$\pm$\,0.07 mag. The peak pseudo-bolometric luminosity of SN~2017gkk is $L_{\rm bol}=(8.37\pm0.19)\times10^{41}\,\mathrm{erg\,s^{-1}}$. Between 25 and 80 days (after maximum), SN~2017gkk displays a fast decline rate of $2.35\pm0.06~\mathrm{mag\times(100d)^{-1}}$ in $V$-band. The photospheric phase spectra are dominated by an H$\alpha$ P~Cygni profile (up to $\sim$\,69~days), which gradually faded and disappeared $\sim$\,90 days after the $V$-band maximum. The \ion{He}{1} lines subsequently emerged after $\sim$\,20 days and became dominant during the transition phase. The nebular phase spectra of SN~2017gkk indicate a relatively low-mass progenitor with $M_{\mathrm{preSN}}\gtrsim3\,M_\odot$ and $M_{\mathrm{ZAMS}}\approx\,12\,\mathrm{M_\odot}$, and it is likely in a binary system. The analytical modeling of the light curve infers a low ejecta mass of $M_{\mathrm{ej}}=1.38^{+0.48}_{-0.31}\, M_\odot$ with relatively lower explosion energy $E_\mathrm{KE}=0.69_{-0.17}^{+0.25}\times10^{51}\,\mathrm{erg}$ and a synthesized $^{56}$Ni mass of $M_{\mathrm{Ni}}=0.076_{-0.024}^{+0.032}\, M_\odot$.
\end{abstract}

\section{Introduction}\label{intro}

The stripped-envelope supernovae \citep[SE-SNe,][]{1941minkowski, Clocchiatti1997ApJ} constitute a subgroup among the core-collapse supernovae (CC-SNe). These events emerge when the stellar core of massive stars ($\gtrsim$10 M$_\odot$) gravitationally collapses at the end point of their evolution \citep{Heger2003H, Woosely-1986W}. The post-explosion observational features of SE-SNe make them distinguishable, particularly in terms of hydrogen lines. These spectral lines are dominantly present in Type II SNe several months after the explosion, but lack in SE-SNe \citep[Type Ib, Ic, IIb; see][for more details]{1997filippenko, Galyam2017G, Modjaz2019M}.  

The observational signatures (weak presence of hydrogen) displayed in SE-SNe are indicative of strong stripping of the hydrogen envelope of the progenitor before the gravitational collapse of the core. A couple of mechanisms have been proposed to explain this stripping from the progenitor star. In a single massive star scenario, stellar wind, eruptive mass loss, or a combination of both may play a significant role \citep{1975MSRSL...9..193C, Woosley-1995, 2006smith, Crowther-2007, 2008A&ARv..16..209P, Smith-2014, Yoon-2015AA}. The binary interaction in intermediate-mass stars is a favored scenario where mass transfer occurs, and envelope mass is removed from the massive to the lower-mass companion star \citep{1984nomoto, 1987nomoto, 1992ApJ...391..246P, 1995woosley, 2002pols, 2004pods, 2010ApJ...725..940Y}. 
Among SE-SNe, IIb events display intermediate observational properties between hydrogen-rich Type II and hydrogen-poor Type Ib. The Balmer lines are present in their early spectra, but disappear progressively with the domination of helium lines at later epochs. This signifies that a small amount of hydrogen \citep[$\sim$0.001--1 M$_\odot$, see][]{Sravan-2019S, Yoon-2017Y} is retained at the point of explosion.

Type IIb transitional spectroscopic features were first noticed in SN~1987K \citep{Filippenko1988F}. Furthermore, a remarkable double-peak light curves has been observed in the majority of Type IIb SNe, such as in SN~1993J \citep{1994richmond, Barbon1995B}, SN~2011fu \citep{SN2011fu_Kumar2013MNRAS.431..308K}, SN~2013df \citep{2014morales, VanDyk2014V}, SN~2016gkg \citep{Tartaglia-2017ApJ...836L..12T, Bersten2018B}, SN~2024iss \citep{Yamanak2025PASJY, Chen-2026C}, SN~2024uwq \citep{Subrayan2025S} and SN~2025aico \citep{SN2025aico-Zhao}. However, in some cases, this feature is weak or absent, like in SN~2008ax \citep{SN2008ax_Pastorello2008MNRAS.389..955P, Roming-2009R}, SN~2022crv \citep{Gangopadhyay2023G, Dong2024D}, SN~2024abfo \citep{SN2024abfo_Reguitti2025A&A...698A.129R} and SN~2022ngb \citep{SN2022ngb_Zhao2025A&A}. Furthermore, a few events exhibited faster light curve evolution relative to other SE-SNe: SN~2022acat \citep{SN2020acat_Medler2022MNRAS.513.5540M} and SN~2024aecx \citep{Zou-2026-ApJ, Xi-2026ApJ...998...98X}.
The heterogeneity in the observational properties of these events hints towards distinct explosion channels, and likely, there are two specific groups of their progenitors. \citet{Chevalier2010C} proposed the extended and compact progenitor scenarios: those exhibiting shock-cooling features immediately after the explosion (eIIb) and those without such characteristics (cIIb).

The high-resolution deep archival imaging of the SN explosion sites acquired with the Hubble Space Telescope (HST) has significantly contributed to constraining the progenitor properties. Various studies inferred that SNe IIb progenitors belong to supergiant (B, K, and Y Type) stars \citep[][and references therein]{Maund2004M, VanDyk2014V, Kilpatrick-2017K, Folatelli2014F, SN2024abfo_Reguitti2025A&A...698A.129R, Niu2024ApJ...970L...9N}. The suggested initial mass has a range of $\sim$\,10--17 $M_\odot$. The analytical modeling \citep[e.g.,][]{Falk1977F, Bersten2012B, 2014nakar, Piro2015P, Sapir2017S, Dessart2018-612A..61D} of the light curves also support these findings. However, in a few cases, WR-like progenitors were also argued \citep{Crockett-2008C, Folatelli-2015F, Gal-Yam-2014G}. Notably, in both cases, the progenitor's origin in a binary system is a favored concept.
Recent large sample-based studies explored the observational properties of IIb events and found diversity in the different explosion parameters \citep{2016MNRAS.457..328L, Taddia2018A&A...609A.136T, 2019Shivvers, 2019MNRAS.485.1559P, Ayala2025A}. It is interesting to study additional events for a better understanding of SE-SNe. Here, we investigate the photometric and spectroscopic properties of SN~2017gkk.

The discovery of SN~2017gkk was reported in galaxy NGC~2748 by the MASTER Global Robotic Network telescope group \citep{Lipunov-2010L} on 2017 August 19.96 UT (MJD~57984.00). The transient was given the internal name MASTER OT J091344.71+762842.4, and its unfiltered discovery magnitude was 15.6 mag \citep{Balanutsa2017B}. Gaia last non-detection at the SN location was on 2017 August 06.99 (MJD~57971.98) at 21.5 mag ($G$-Gaia band). The photometric observations,  from the Konkoly Observatory, Hungary, obtained between 2017 August 25 and 2017 August 29, confirmed a rapid brightening of the SN \citep{Vinko-2017V}. Later, the SN discovery was also confirmed by Koichi Itagaki on 2017 August 31.79 UT (MJD~57996.79) with a clear filter magnitude of 14.7 \citep{Itagaki-2017I}. The spectrum of SN~2017gkk, taken on 2017 September 02.24 with the ALFOSC instrument mounted on the 2.56 m Nordic Optical Telescope, classified it as a Type IIb SN \citep{Onori-2017O}.

The properties of SN~2017gkk have been preliminarily explored in two recent studies. \citet{Koivisto-17gkk-thesis} examined its light curve and spectroscopic evolution, classifying it as a fast-evolving, faint, and relatively low energetic Type IIb SN. Their evaluated ejecta and $^{56}$Ni masses were $\sim2.4 M_\odot$ and $\sim0.2M_\odot$, respectively. Based on [\ion{Ca}{2}]/[\ion{O}{1}] ratio, the progenitor was estimated as a $\sim23M_\odot$ star. They argued the similarity of SN~2017gkk with SNe 1993J and 2008ax but suggested further analysis for a better constraint of observational and progenitor properties. In a subsequent study, \cite{Niu2024ApJ...970L...9N} identified its progenitor in the pre-explosion HST images. Through SED fitting and modeling of its evolutionary path, they concluded that SN~2017gkk arose from a dusty yellow supergiant, with an initial mass ($M_{\mathrm{ZAMS}}$) of $\sim$ $15\,M_{\odot}$, surrounded by a thin shell of circumstellar material (CSM).

In this paper, the observations and data analysis are described in Section~\ref{obs}. The photometric properties, including light curve evolution, extinction correction, absolute light curve, and pseudo-bolometric light curve, are presented in Section~\ref{sec:pho}. The spectroscopic properties are presented in Section~\ref{sec:spec}. In Section~\ref{discussion}, the light curve modeling and nature of the progenitor are discussed. Finally, the results are summarized in Section~\ref{sec:summary}.

\begin{figure}
\centering
\includegraphics[width=\columnwidth]{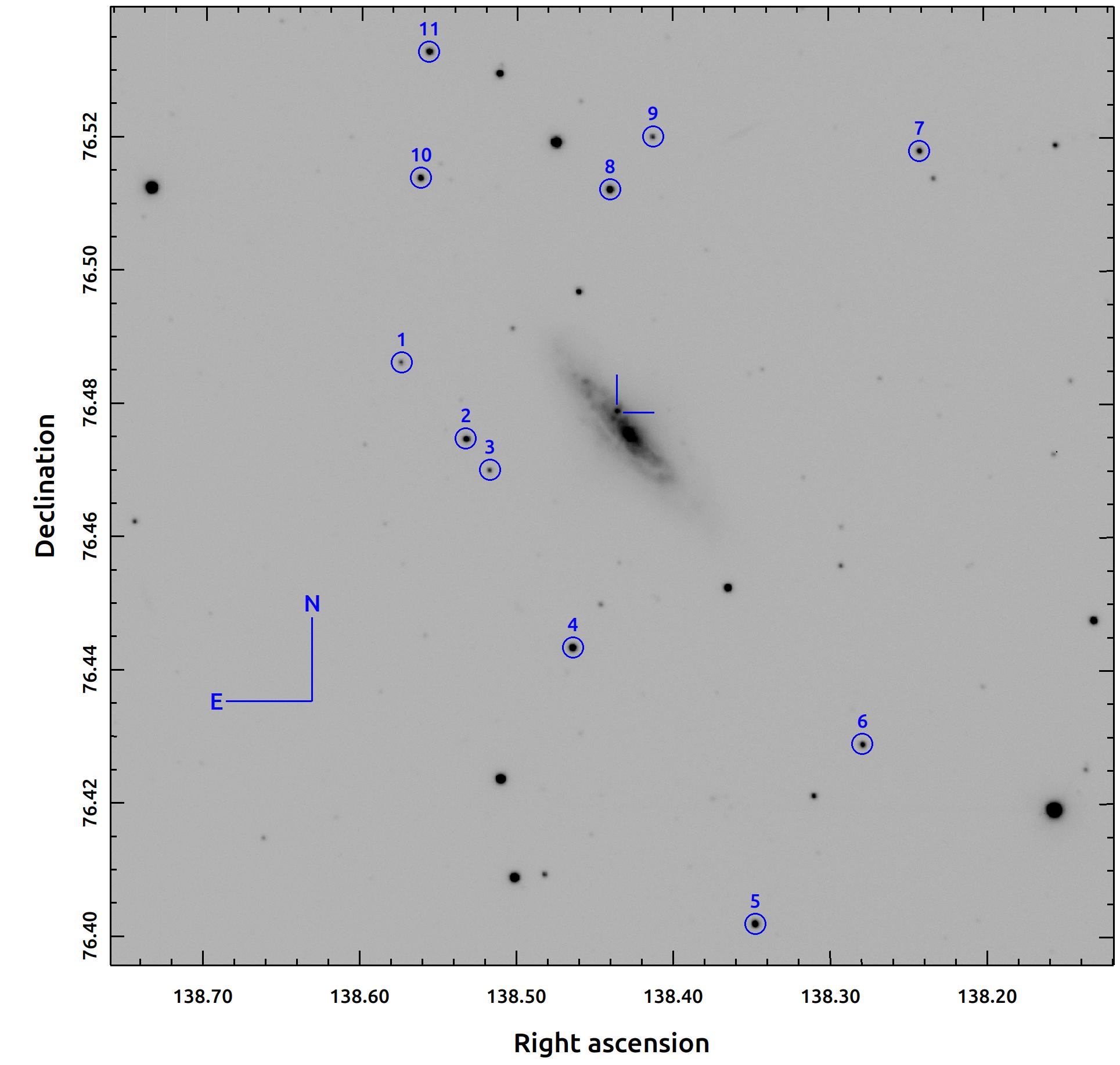}
\caption{Identification chart of SN~2017gkk (HCT $R$-band image taken on 2017 September 27, see Section~\ref{obs_ph}). The secondary standard stars are indicated with IDs 1--11, and the SN location is marked with a crosshair symbol.}
\label{fig:sn-field}
\end{figure}

\section{Observation and data analysis}\label{obs}

SN~2017gkk (RA=$9^{h}13^{m}44^{s}.570$, DEC=$+76^\circ28^{\prime}44.54^{\prime\prime}$; ICRS) is located in the host galaxy NGC 2748, at $5^{\prime\prime}.38$ east and $13^{\prime\prime}.31$ north from the center of the host galaxy (see Figure~\ref{fig:sn-field}). The redshift of the host galaxy is obtained from the NASA/IPAC Extragalactic Database (NED)\footnote{\url{https://ned.ipac.caltech.edu}}, which gives $z=0.004923\pm0.000007$. We adopt the distance of $D_{\mathrm{L}}=19.05\pm0.59\,\mathrm{Mpc}$ from the NED, which is the average distance of various measurements by the Tully-Fisher relation \citep{McGaugh2000ApJ...533L..99M}. The corresponding distance modulus is $\mu_{\mathrm{L}} = 31.40 \pm 0.07\,\mathrm{mag}$. 

\begin{table}[htbp]
\scriptsize
\centering
\caption{Details of SN~2017gkk and its host galaxy NGC~2748.}\label{tab:basicinfo}
\begin{tabular}{llc}
\hline
Parameters                         &   Value                            &  Reference    \\ 
\hline
SN~2017gkk:                        &                                    &               \\    
\quad RA (J2000)                   &   $9^{h}13^{m}44^{s}.57$           & 1             \\   
\quad DEC (J2000)                  &   $+76^\circ28^{\prime}44^{\prime\prime}.54$& 1    \\ 
\quad Discovery date (UT)          &   2017 August 19.96                & 2             \\
                                   &   (MJD~57984.00)                   &               \\
\quad $E(B-V)_{total}$             &  0.45 $\pm$ 0.16 mag          & Section~\ref{ext}  \\ 

NGC~2748:                          &                                    &               \\
\quad Type                         &   SAbc                             & 3   \\
\quad Redshift                     &   $0.004923\pm0.000007$            & 3   \\
\quad Distance                     &   $19.048 \pm 0.593$ Mpc           & 3   \\
\quad Distance modulus             &   $31.40 \pm 0.07$ mag             & 3   \\ 
\hline
\end{tabular}  
\begin{minipage}{\textwidth}
$^{1}$ \url{https://www.wis-tns.org/object/2017gkk} \\
$^{2}$ \url{https://www.astronomerstelegram.org/?read=10667}\\
$^{3}$ NASA/IPAC Extragalactic Database (NED) \\
\end{minipage}
\end{table}

\subsection{Photometric observation}\label{obs_ph}

The photometric monitoring of SN~2017gkk was carried out in  Bessell's $BVRI$ bands, between 2017 September 03 (MJD~57999.9) and 2018 April 07 (MJD~58215.7), using the Himalayan Faint Object Spectrograph Camera (HFOSC) mounted on the $f/9$ Cassegrain focus of the 2-m Himalayan Chandra Telescope (HCT) of the Indian Astronomical Observatory, Hanle, India \citep{2014Prabhu}.  The HFOSC instrument is equipped with a liquid nitrogen-cooled 2K $\times$ 4K pixels SITe CCD chip with pixel size 15 $\times$ 15 $\mu$m (gain: 1.22 e$^{-}$/ADU and readout noise: 4.87 e$^{-}$). The central 2K $\times$ 2K region used for imaging covers a field of 10$\arcmin$\,$\times$\,10$\arcmin$ on the sky (plate scale: 0.296 arcseconds per pixel).

The raw science images were preprocessed with the help of bias and twilight flat frames obtained on the respective observing nights. Cosmic rays were then removed. The standard tasks available in the data reduction software \texttt{IRAF}\footnote{\texttt {IRAF} stands for Image Reduction and Analysis Facility. It is distributed by the National Optical Astronomy Observatory, which is operated by the Association of Universities for Research in Astronomy (AURA) under a cooperative agreement with the National Science Foundation.} were utilized for the image preprocessing. Multiple science frames in the respective bands were taken on some nights to achieve a better signal-to-noise ratio. The cleaned images were co-added after the geometrical alignment.

The secondary standard stars in the SN field were calibrated by observing the Landolt photometric standard fields \citep{1992landolt} PG~2331+055, PG~2213--006, and PG~2213--006 on 2017 September 27 and 2017 October 01 under good sky conditions. The standard stars have a brightness and color range of $12.71\le V \le 15.18$ mag and $-0.22 \le B-V \le 0.82$ mag, respectively. In the calibration, the average atmospheric extinction of the site was adopted from \citet{Stalin2008}.
The photometric zero-points were estimated using the average colour terms for the telescope detector system, and these were utilized to calibrate the secondary standards in the SN field. A set of eleven such stars are shown in Figure~\ref{fig:sn-field} and their calibrated $BVRI$ magnitudes are provided in Table~\ref{apptab:calibphotometryinfo}.

As can be seen in Figure~\ref{fig:sn-field}, the SN is deeply embedded inside the host galaxy, and therefore, template subtraction is necessary to obtain the true flux of the SN. The template images of the SN field were taken with HCT on 2018 November 08 ($\sim$433 days after discovery) in good sky conditions. The previously acquired science frames were aligned with respect to the template images, followed by PSF matching in respective bands, and finally, image subtraction was performed. The template-subtracted images from different epochs were used to extract the SN instrumental magnitudes. The nightly zero points were determined using the secondary standards in the field, and the SN magnitudes were calibrated differentially. The photometric errors include the \textit{IRAF} resulted errors in aperture photometry and uncertainties in the nightly zero points, added in quadrature. The calibrated SN magnitudes in $BVRI$-bands and the associated errors are listed in Table~\ref{apptab:photometrydata}.

In addition to the HCT photometry, we also used four epochs (between MJD 57990.79 to 57994.79) of $BVRI$-band archival data from \citet{Vinko-2017V} in our study. These photometric data were prior to the HCT observations and are available online\footnote{https://www.astronomerstelegram.org/?read=10686}.

\subsection{Spectroscopic observation}\label{obs_sp}
Low-resolution optical spectroscopic observations of SN~2017gkk were obtained at 17 epochs during 2017 September 03 (MJD~57999.83) to 2017 December 29 (MJD~58116.65). Two grisms Gr\#7 (3500--7800 \AA) and Gr\#8 (5200--9250 \AA), available with HFOSC, were used to obtain the SN~2017gkk spectra. A log of these observations is provided in Table~\ref{apptab:spectroscopy_log}.

The spectra of the arc lamp and spectrophotometric standards were obtained along with the SN  for calibration purposes. The spectroscopic data reduction was carried out in a standard way using {\small IRAF}. After correcting the observed frames for bias and flat fields, one-dimensional spectra were extracted using the optimal extraction method. 

The wavelength calibration was done by applying the dispersion solutions obtained using arc lamp spectra. The night-sky emission lines were utilized to secure the wavelength calibration, and small shifts were applied wherever necessary. The spectrophotometric standard observed on the same night was used to obtain the instrumental response curve for flux calibration of the spectra. For those nights where standard star observations were not possible, the response curves obtained during nearby nights were used. To construct a single flux-calibrated spectrum, the spectra in both grisms (Gr\#7, Gr\#8) were combined. The spectra were then scaled with respect to the calibrated photometric fluxes to bring them to an absolute flux scale. Finally, the SN spectra were corrected for the host galaxy redshift of $z$ = 0.006191 (from NED).

\begin{figure}
\centering
\includegraphics[width=\columnwidth]{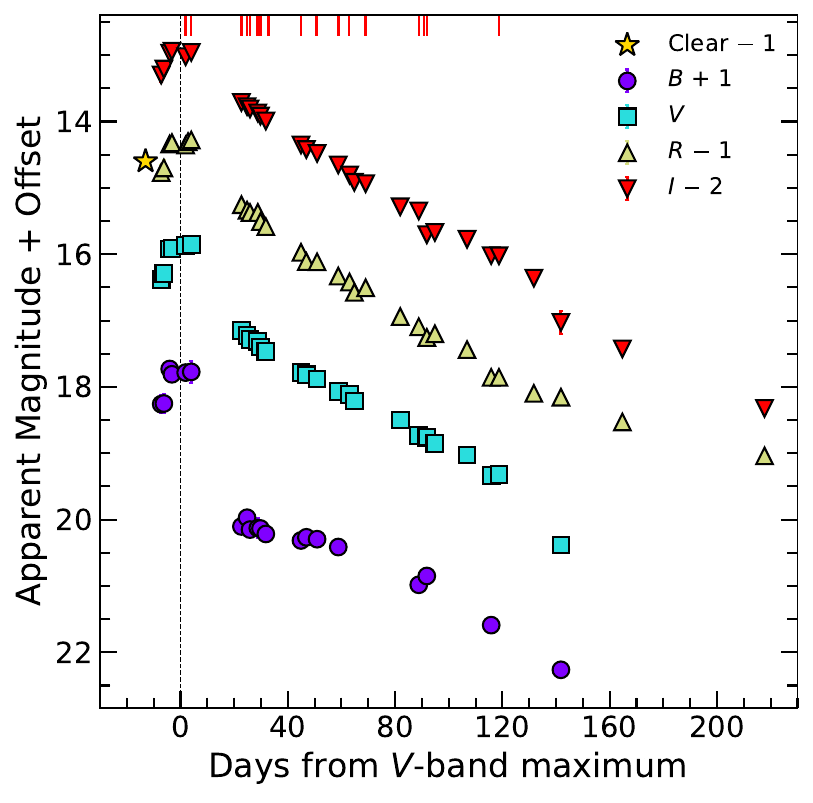}
\caption{$BVRI$-band light curve evolution of SN~2017gkk, showing apparent magnitudes with arbitrary offsets. Error bars are included for all photometric points, but they are generally smaller than the marker size. A star symbol indicates the discovery magnitude in the clear band \citep{Balanutsa2017B}. A dotted vertical line indicates the epoch of $V$-band maximum (see Section~\ref{sec:pho:applc}). The small vertical lines (red) on the top mark the epochs of spectroscopic observations used in this study.}
\label{fig:apparent_lc}
\end{figure}

\section{Photometric evolution}\label{sec:pho}
\subsection{Apparent light curve}\label{sec:pho:applc}

The optical multi-band light curve of SN~2017gkk is presented in Figure~\ref{fig:apparent_lc}. This is constructed using our photometric and publicly available data \citep{Vinko-2017V} as listed in Table~\ref{apptab:photometrydata}. The discovery magnitude from \citet{Balanutsa2017B} is also shown with a star symbol. Our observations span from MJD~57990 to MJD~58215, a period of $\sim$5 months after discovery, tracking evolution from peak to nebular phase. As SN~2017gkk was discovered close to its maximum light (at MJD~57997), the crucial early rising phase is missing from our dataset. Furthermore, the coverage during the post-peak decline is also relatively sparse.

Considering the wide interval ($\sim$40 days) between the last non-detection and the first detection from GAIA, the explosion epoch of SN~2017gkk is poorly constrained by direct observations. To estimate it, we determined the zero-flux epoch in each photometric band by extrapolating the early-time light curves and then took the average. For this purpose, we employed the data-driven artificial neural network code \textsc{refann}\footnote{\url{https://refann.readthedocs.io}} \citep{Wang2020ApJS..246...13W} to fit the photometric data and extrapolate to the zero-flux time, while also estimating the associated uncertainties. The derived zero-flux epochs are MJD~$57985.80^{+0.94}_{-1.03}$ in the $B$-band, MJD~$57985.12^{+0.68}_{-0.71}$ in the $V$-band, MJD~$57984.70^{+0.96}_{-1.06}$ in the $R$-band, and MJD~$57982.71^{+1.21}_{-1.34}$ in the $I$-band. Averaging these values yields a final estimated explosion epoch of MJD~$57984.89^{+0.44}_{-0.48}$.
\citet{Koivisto-17gkk-thesis} suggested the explosion epoch of SN~2017gkk as MJD~57982 by comparing its bolometric light curves with SNe 1993J and 2008ax. Our estimation is consistent with this study, with an uncertainty of $\sim$3 days. However, due to the inherent uncertainties associated with extrapolating from data with sparse early-phase coverage and the complicated mechanics of the early phase, this value should be treated with caution. Further, it is not trustworthy to state about the possible shock-cooling phase based on the single epoch discovery magnitude shown in Figure~\ref{fig:apparent_lc}. Consequently, in our subsequent model fitting, we will treat the explosion epoch as a lower limit only.

The photometric data, supplemented with archival data from \citep{Vinko-2017V}, provide a relatively good coverage around the peak in each band. To determine the peak epoch and magnitude, we employed \textsc{refann} to obtain a smooth, continuous representation of the light curve and applied a Monte-Carlo polynomial fitting to estimate the peak parameters and their uncertainties. The estimated epoch of light curve maximum are MJD 57997.30$\pm$1.25 in the $B$-band, MJD 57997.98$\pm$0.56 in the $V$-band, MJD 57997.62$\pm$0.56 in the $R$-band, and MJD 57996.89$\pm$0.62 in the $I$-band. The rise times and peak magnitudes for all bands are summarized in Table~\ref{apptab:apparentlc}. The $13.09_{-0.74}^{+0.71}$ days rise time in $V$-band of SN~2017gkk is notably shorter as compared to other well-studied type-IIb supernovae, as SN~2011hs (16.5 days; \citealt{SN2011hs_Bufano2014MNRAS.439.1807B}), SN~2011ei (17.0 days; \citealt{SN2011ei_Milisavljevic2013ApJ...767...71M}), SN~2020acat (16.5 days; \citealt{SN2020acat_Medler2022MNRAS.513.5540M}), and SN~2024abfo (22.1 days; \citealt{SN2024abfo_Reguitti2025A&A...698A.129R}).

The post-peak photometric coverage for SN~2017gkk is sparse within the initial $\sim$25 days but has better sampling at later epochs ($\gtrsim$25 days). To quantify the evolution at later phases, we performed linear fits to the light curves between 25--80 days post $V$-band maximum. The decline rates of each bands are $\gamma_{25<t<80}(B)=0.73\pm0.10$, $\gamma_{25<t<80}(V)=2.35\pm0.06$, $\gamma_{25<t<80}(R)=2.78\pm0.11$, and $\gamma_{25<t<80}(I)=2.62\pm0.07\,\mathrm{mag\times(100\,days)^{-1}}$. The faster decay rate in the redder bands is characteristic of the photosphere receding through the ejecta due to recombination in the outer, cooler layers \citep{Nagy2014}, while the inner ejecta, powered by $^{56}$Co decay, becomes progressively transparent \citep{SN2011fu_Kumar2013MNRAS.431..308K}. The decline rates of SN~2017gkk are presented in Table~\ref{apptab:apparentlc}.

\subsection{Extinction correction}\label{ext}

A reliable determination of the line of sight extinction towards the SN direction is crucial, as it significantly affects the observational parameter estimates. The reddening may arise due to the dust from the Milky Way (MW) and the host. The MW extinction is typically corrected following the dust maps \citep{Schlegel98, Schlafly11, Schlafly-2016S} and various reddening laws \citep{1989cardelli, 1999fitzpatrick}. It is more challenging for an accurate estimation of the host reddening in the absence of explicit information on its dust properties. Some of the most widely adopted observational techniques (though with caveats) include the intrinsic color comparison \citep{Drout2011D, Stritzinger2018S}, \ion{Na}{1} D equivalent width \citep{Turatto2003fthp.conf..200T, Rodriguez-2023R}, and the Balmer decrement \citep{Dominguez2013D}.

The MW extinction towards SN~2017gkk, adopted from \cite{Schlafly11} and NASA/IPAC Infrared Science Archive\footnote{\url{https://irsa.ipac.caltech.edu}} dust extinction map, is $E(B-V)_{\mathrm{MW}} = 0.0232 \pm 0.0011\,\mathrm{mag}$. Significant host attenuation is expected as the SN is located deep inside it, and so here we examined different approaches to constrain the host extinction. By examining the multiband light curve of Ibc SNe, \citet{Drout2011D} found that $\sim$10 d past $V$-band maximum, these events exhibit very small scatter in the $(V-R){_0}$ color with a mean $(V-R){_0}$ color of 0.26 $\pm$ 0.06 mag. Considering the unavailability of SN~2017gkk data during this phase, we performed interpolation in these bands, and $(V-R)$ was derived as 0.68$\pm$0.04 mag, which finally gives $E(B-V)$=0.70 mag.

We applied the \citet{Rodriguez-2023R} empirical relation for SE-SNe, which relies on the EW of \ion{Na}{1} D. For this purpose, the EW of \ion{Na}{1} D was estimated as $1.457 \pm 0.172\,\mathrm{\AA}$ from the stacked spectrum at three epochs ($+3.96\,\mathrm{days}$, $+22.70\,\mathrm{days}$, $+24.87\,\mathrm{days}$ from $V$-band maximum). The estimated $E(B-V)$ was found to be 0.37 $\pm$ 0.10 mag. The Balmer decrement method \citep{Dominguez2013D} was also implemented to the stacked spectrum obtained on 2017 September 24, 26, and 27. The derived $E(B-V)$ was 0.48 $\pm$ 0.14 using this method.

It is worth noting that the host $E(B-V)$ estimation based on the color-matching method was primarily for the Ibc SNe \citep{Drout2011D}. Furthermore, our estimated value for SN~2017gkk is based on the interpolated magnitudes (cf. sparse data sampling), and therefore, it is likely unreliable. The host extinction values obtained from the remaining two methods are in better agreement; hence, an average of both (0.37 and 0.48 mag) was considered as the reddening arising from the SN host. The total extinction (MW+host) of SN~2017gkk is $E(B-V)_{\mathrm{Total}} = 0.45 \pm 0.16\, \mathrm{mag}$ and this value was considered in our further analysis.

\begin{figure}
\centering
\includegraphics[width=\columnwidth]{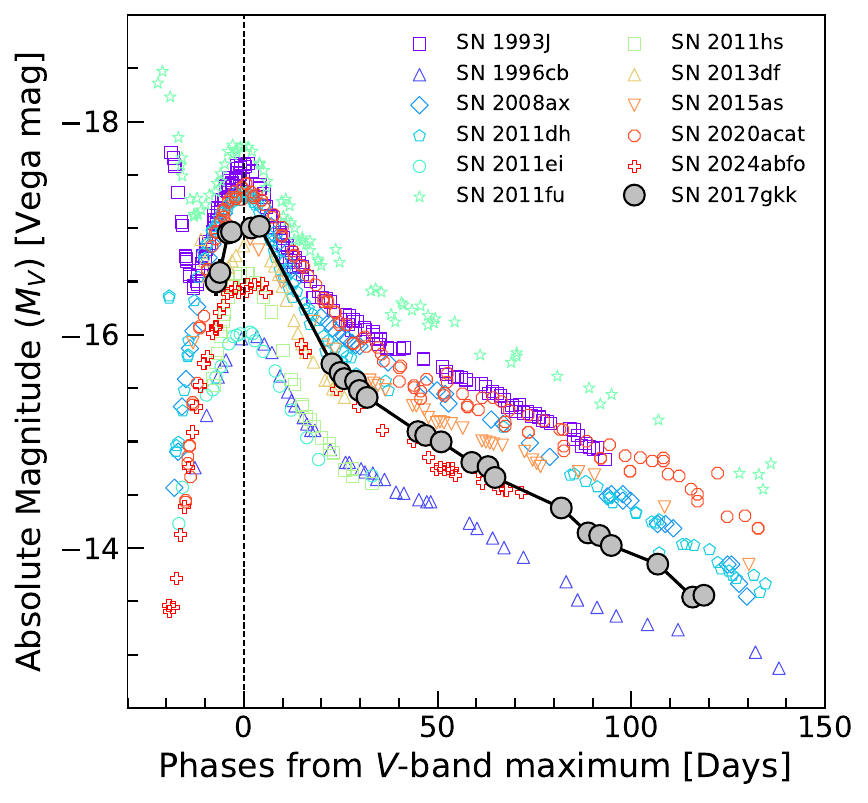}
\caption{$V$-band absolute light curve of SN~2017gkk along with other Type IIb events. All the compared SNe are corrected for reddening, and distance measurements were adopted from the respective studies as mentioned in Section~\ref{fig:absolute_lc}.}
\label{fig:absolute_lc}
\end{figure}

\begin{figure}
\centering
\includegraphics[width=\columnwidth]{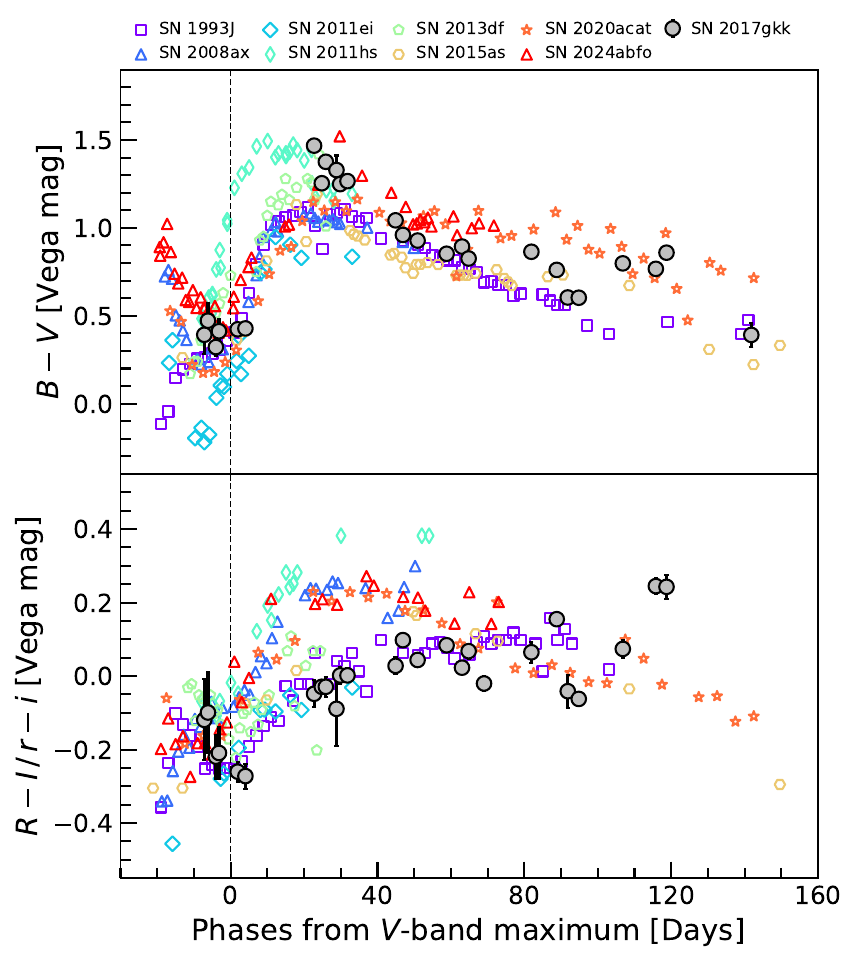}
\caption{$B-V$ and $R-I\,/\,r-i$ color evolution of SN~2017gkk and other type IIb events. The reddening is corrected for all the targets and converted to the Vega system if necessary.}
\label{fig:colorevo}
\end{figure}

\subsection{Absolute light curve and color evolution}
\label{sec:pho:abslc_color}

The $V$-band absolute light curve of SN~2017gkk, along with other Type IIb events, is presented in Figure~\ref{fig:absolute_lc}. We noticed that Type IIb SNe show a variety in their radioactive-powered peak, having a relatively wide range of $\sim$2 mag \citep[see also][]{Taddia2018A&A...609A.136T}. The peak $V$-band absolute magnitude of SN~2017gkk is --17.01 $\pm$ 0.07 mag, which could be considered the bridge between the bright (e.g., SN~1993J and SN~2011fu) and the faint events (e.g., SN~2011ei and SN~2024abfo). This feature suggests that SN~2017gkk should be considered a moderately luminous SN.

The color evolution of SN~2017gkk is presented in Figure~\ref{fig:colorevo} and compared with other IIb SNe. All colors have been corrected for the respective reddening and are expressed in the Vega system. The overall color evolution trend of SN~2017gkk is similar to that of the compared objects. Although we have sparse coverage near the peak, it appears that the $B-V$ color of SN~2017gkk has a red value of $1.49\pm0.02$~mag near the peak, similar to SNe~2011hs and 2024abfo. Following this red peak, it undergoes a rapid blue phase. Interestingly, the $R-I$ color evolution of SN~2017gkk follows that of SN~1993J well up to $\sim$60 days post-maximum, but later it is significantly scattered.

\begin{figure}[htbp]
\centering
\includegraphics[width=\columnwidth]{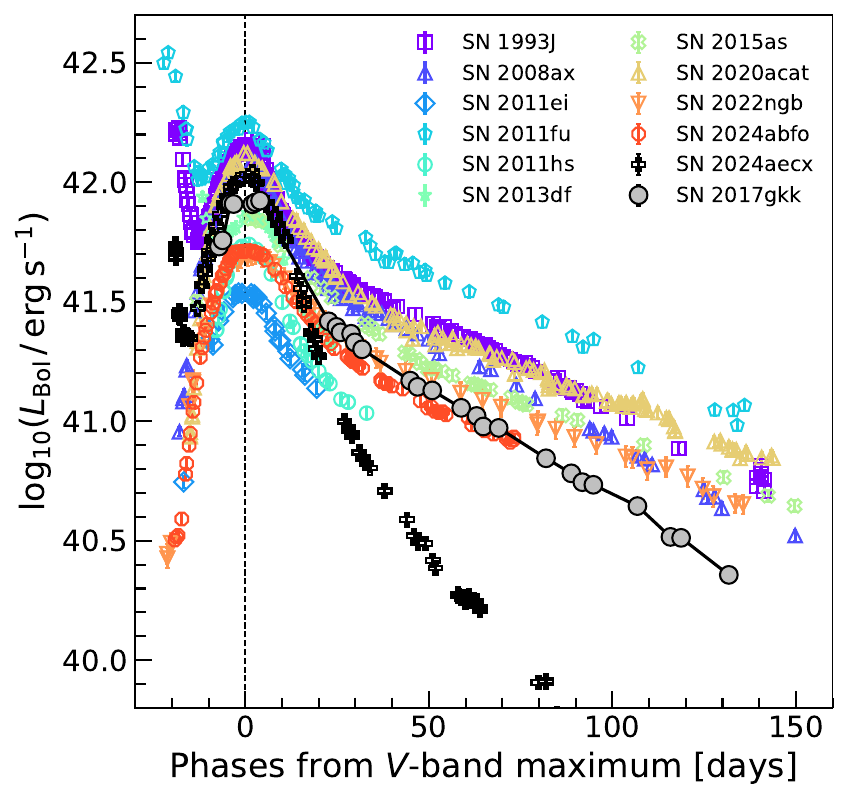}
\caption{Pseudo-bolometric light curve of SN~2017gkk along with other Type IIb SNe for comparison. The pseudo-bolometric light curves are constructed by integrating the fluxes from $B$ to $I$ bands (see Section~\ref{sec:pho:bollc}).}
\label{fig:pseudobollc}
\end{figure}

\begin{figure}[htbp]
\centering
\includegraphics[width=\columnwidth]{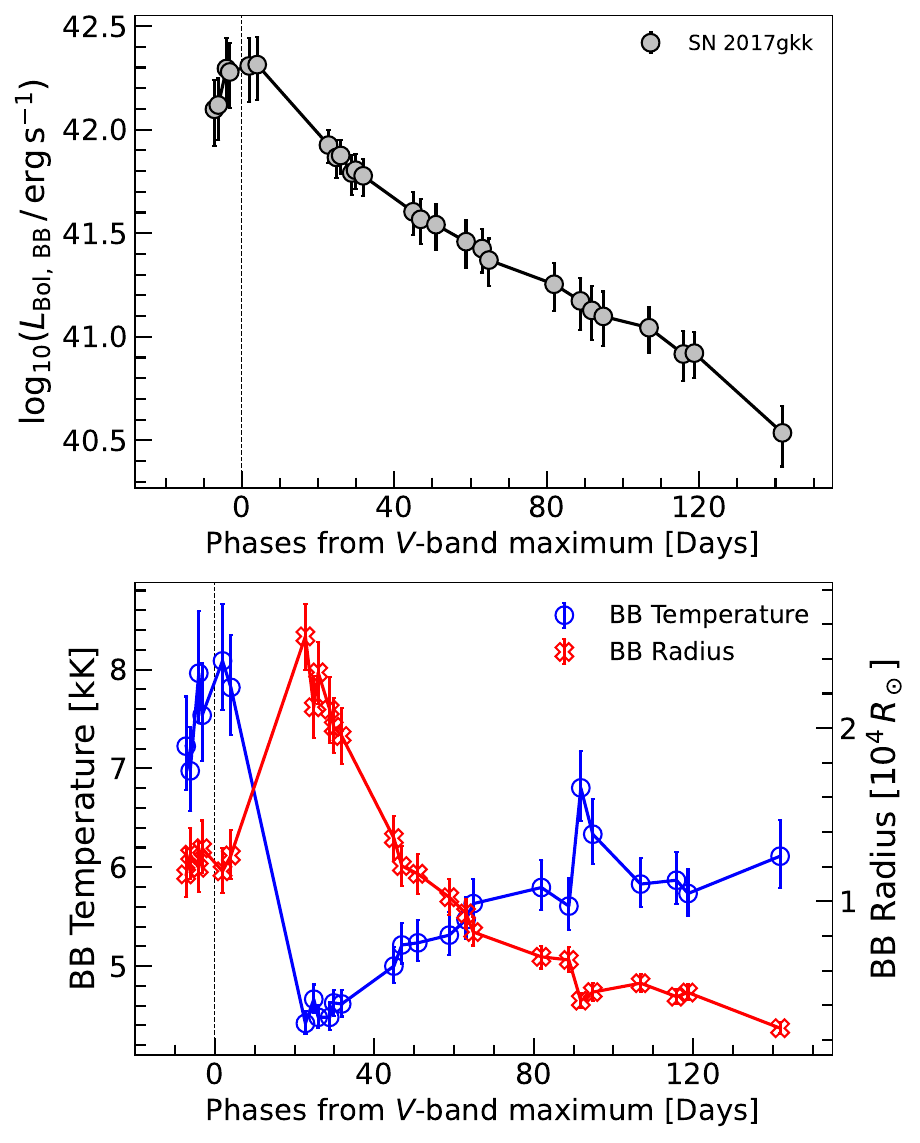}
\caption{Bolometric light curve evolution of SN~2017gkk. \textit{Upper panel}: BB-corrected bolometric light curve of SN~2017gkk. \textit{Lower panel}: BB temperature and radius evolution of SN~2017gkk.}
\label{fig:bollc}
\end{figure}

\subsection{Pseudo-bolometric light curve}
\label{sec:pho:bollc}
The pseudo-bolometric light curves of SN~2017gkk and other well-studied events were constructed uniformly. The photometric data were corrected for reddening, and the resulting magnitudes were converted to flux. The pseudo-bolometric light curves were then generated by integrating these fluxes over the optical wavelength range from the $B$ to the $I$ band using \textsc{superbol} \citep{Nicholl2018}. Figure~\ref{fig:pseudobollc} shows the resulting pseudo-bolometric light curve of SN~2017gkk, plotted along with the objects used for comparison.

The pseudo-bolometric light curve of SN~2017gkk reached a peak luminosity of $L_{\rm bol}=(8.37\pm0.19)\times10^{41}\,\mathrm{erg\,s^{-1}}$ at MJD~$57998.82\pm0.90$. This value is similar to those of intermediate luminous events like SN~2013df and SN~2024aecx, within the ranges of the lower luminosity end and high luminosity events like SN~1993J and SN~2011fu. Furthermore, the light curve of SN~2017gkk declines more rapidly, with its luminosity fading by approximately 0.9 dex per 100 days, in contrast to a rate of $\sim 0.7$ dex per 100 days for the comparison samples. The rapid decline of the bolometric light curve points to a substantial gamma-ray leakage at late times \citep{Clocchiatti1997ApJ}, a phenomenon that can be attributed to either a low ejecta mass or reduced gamma-ray opacity. Nevertheless, SN~2024aecx is the most fast evolving among the compared events \citep[see also][]{Zou-2026-ApJ, Xi-2026ApJ...998...98X}.

We constructed the bolometric light curve by fitting a black-body (BB) function to the spectral energy distribution (SED) at each observational epoch. The fitting was performed using an MCMC algorithm to robustly determine the best-fit posteriors, namely the temperature ($T_{\mathrm{BB}}$) and radius ($R_{\mathrm{BB}}$), along with their associated uncertainties. The bolometric luminosity at each epoch was subsequently derived using the Stefan-Boltzmann law, $L_{\mathrm{Bol, BB}} = 4\pi R_{\mathrm{BB}}^2\sigma T^{4}_{\mathrm{BB}}$. Figure~\ref{fig:bollc} presents the resulting light curve and the evolution of the best-fit BB temperature and radius.

The peak bolometric luminosity of SN~2017gkk is $L_{\rm bol,\,BB}=(2.15\pm0.05)\times10^{42}\,\mathrm{erg\,s^{-1}}$, reached on MJD~$58000.60\pm1.17$. This luminosity falls within the range of  $L_{\mathrm{Bol,\,BB}} = 5$--$50\times10^{41} \mathrm{erg\, s^{-1}}$ observed in the large sample of SE-SNe analyzed by \cite{Taddia2018A&A...609A.136T}. By applying the method described in Sect.~\ref{sec:pho:applc} to the bolometric light curve, the explosion epoch and bolometric rise time of SN~2017gkk are estimated as MJD~$57984.43_{-0.14}^{+0.10}$ and $\sim16.17$ days, respectively, confirming its rapid evolution.

The evolution of the BB temperature and photospheric radius of SN~2017gkk follows the typical trend observed in Type IIb SNe. Before reaching its peak luminosity, the temperature rises from $6.97_{-0.41}^{+0.44}\,\mathrm{kK}$ to $8.09_{-0.50}^{+0.58}\,\mathrm{kK}$, while the photospheric radius remains nearly constant at $\sim12000\,R_\odot$, consistent with a rapidly increasing luminosity. Following the peak, the ejecta undergoes rapid adiabatic expansion, causing the temperature to drop sharply to $\sim4.6\,\mathrm{kK}$ and the radius to expand to $\sim25500\,R_\odot$. At later phases, the temperature begins to rise again slowly as the photosphere recedes, revealing the hotter, inner layers of the ejecta as they become transparent. While this evolutionary pattern is consistent with the sample presented by \cite{Taddia2018A&A...609A.136T}, the temperature of SN~2017gkk lies at the lower end of their distribution.

\begin{figure*}
\centering
\includegraphics[width=\textwidth]{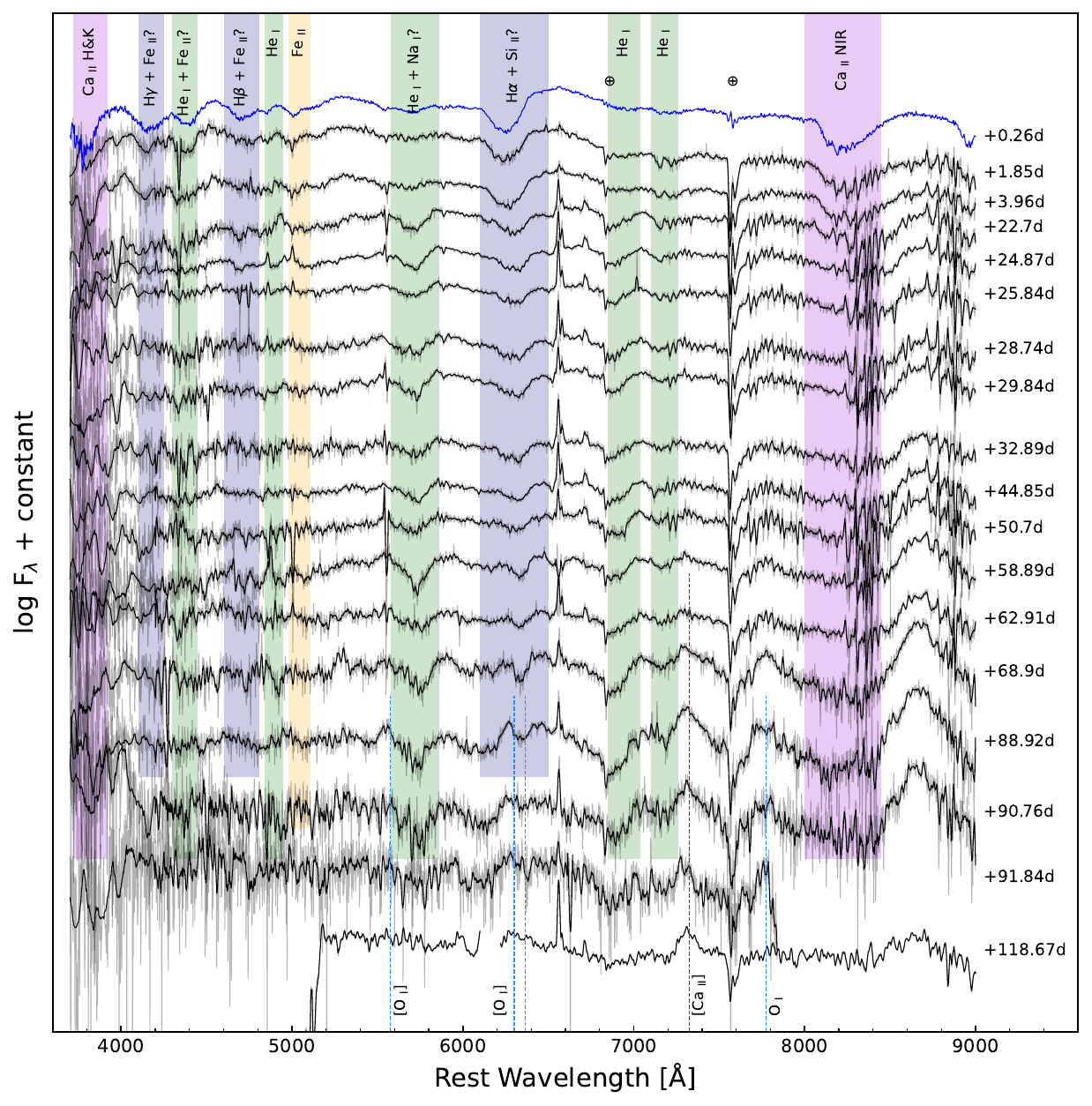}
\caption{The spectroscopic evolution of SN~2017gkk. Prominent spectral lines are marked either by the color region in the P~Cygni absorption or by dashed lines at the rest wavelength. The marked phases are relative to the $V$-band maximum. The spectra are redshift-corrected, and vertical shifts have been applied for clarity. The telluric lines are indicated with encircled plus symbols. All the spectra from HFOSC/HCT were smoothed, and the original spectra are presented in gray. Savitzky–Golay filter with a window length of 21 and a polynomial order of 3 was applied to smooth the spectra. The early spectrum at $\sim$0.3 day was retrieved from NOT/ALFOSC\footnote{https://www.wis-tns.org/object/2017gkk/classification-cert} and plotted in blue on top.}
\label{fig:specevo}
\end{figure*}

\section{Spectroscopy}\label{sec:spec}
\subsection{Spectral evolution}\label{sec:spec:evo}

The spectral evolution of SN~2017gkk is presented in Figure~\ref{fig:specevo}. The spectroscopic observations span from near peak luminosity ($+$0.26 days relative to V-band maximum) to the early nebular phase (+118.67 days).
Narrow emission lines in the spectra, such as H$\alpha$, H$\beta$, [\ion{N}{2}]~$\mathrm{\lambda\lambda}$6548,6583~\AA\ and [\ion{S}{2}]~$\mathrm{\lambda\lambda}$6716,6731~\AA\ are attributed to an H II region in the host galaxy. The  \ion{Na}{1}~D lines at $+$1.85~days, $+$3.96~days, and $+$22.7~days suggest that this event is reddened significantly, consistent with its location \citep{Turatto2003fthp.conf..200T, Poznanski2012MNRAS.426.1465P}. In the early phase ($+$0.26 to $+$25.8 days), the spectra show a blackbody continuum that progressively becomes redder. 

In the early and transition phases, the spectra are dominated by several prominent P-Cygni profiles, which represent the optically thick, expanding ejecta. The early-phase spectra ($+$0.26~days to $+$32.89~days) show prominent H$\alpha$, H$\beta$, and weak \ion{He}{1}~$\mathrm{\lambda}$5876~\AA\ line, \ion{Si}{2} lines. The \ion{He}{1}~$\mathrm{\lambda}$5876~\AA\ line becomes increasingly prominent as the SN evolves and the hydrogen features fade, a characteristic spectral evolution for Type IIb SNe \citep{Filippenko1988F}. Additionally, the \ion{Si}{2}~$\mathrm{\lambda}$6355~\AA\ line gradually emerges on the blue wing of the H$\alpha$ profile, a common feature in SE-SNe \citep{SN2020cpg_Medler2021MNRAS.506.1832M}. At bluer wavelengths, lines of metals such as \ion{Fe}{2}, \ion{Sc}{2}, and \ion{Ti}{2} appear and grow in prominence over time. This suggests that the photosphere recedes deeper into the ejecta, gradually revealing the inner, metal-rich layers.
By day $+$88.92~days, the H$\alpha$ disappears, and the spectrum is dominated by \ion{He}{1}, indicating a transition to a type Ib supernova. During  $+$90.76 to $+$118.67 days, the forbidden emission lines due to [\ion{Ca}{2}]~$\mathrm{\lambda\lambda}$7291, 7324~\AA\ and [\ion{O}{1}]~$\mathrm{\lambda\lambda}$6300, 6364~\AA \ emerged, indicating  evolution toward the nebular phase.

\subsection{Identification of lines}\label{sec:spec:synapps}

\begin{figure*}[htbp]
\centering
\includegraphics[scale=0.8]{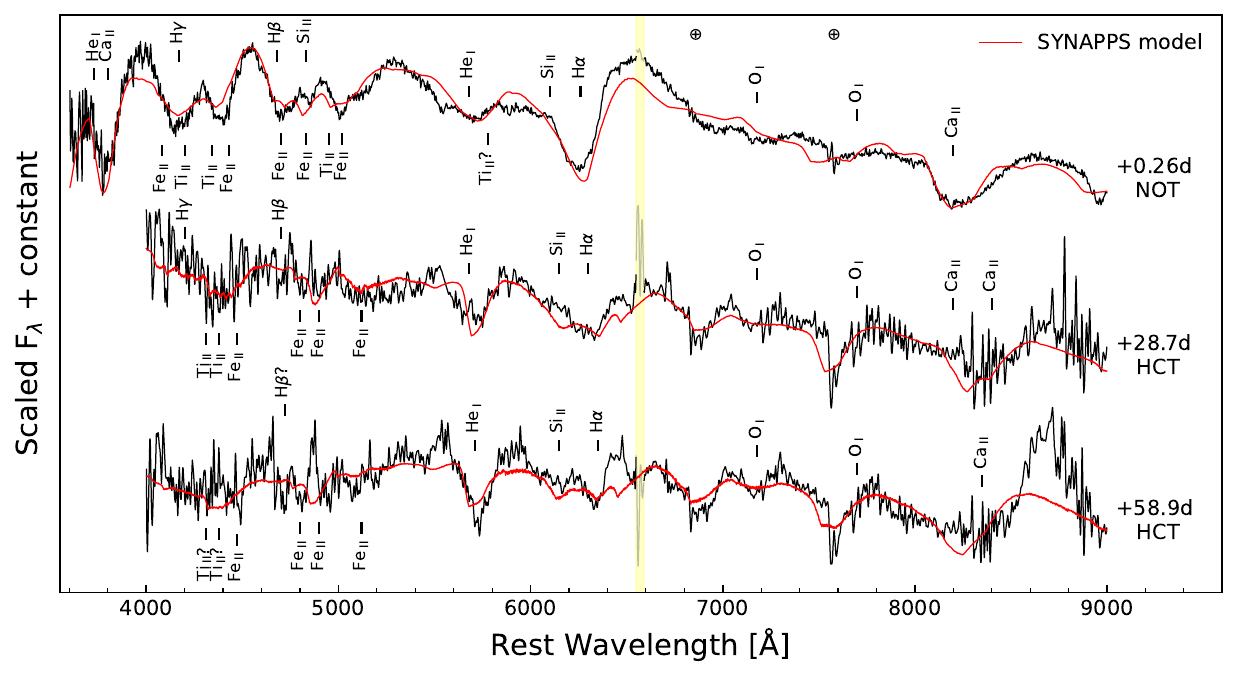}
\caption{Different phase spectra of SN~2017gkk with best-fit model from \textsc{SYNAPPS}. The black and red lines represent the observational data of SN~2017gkk and the best-fit model, respectively. Lines are marked with the short black line in the P~Cygni absorption feature. The shaded region (orange) represents the narrow lines from the \ion{H}{2} region of the host galaxy. The telluric lines are indicated with encircled plus symbols.}
\label{fig:specid}
\end{figure*}

To have a robust identification of the spectral lines in our data, we modeled the spectra using the \textsc{SYNAPPS} and \textsc{SYN++} codes \citep{Thomas2011PASP..123..237T} at three epochs: an early spectrum at $+$0.26~days, a peak-light spectrum at $+$29.84~days, and a late-transition spectrum at $+$58.89~days. The model assumes that photons escape from a photosphere, the location of which is determined by the line optical depth. The results of these fits are presented in Figure~\ref{fig:specid}. For the fitting routine, we included ions corresponding to lines that are prominent in our spectral evolution, as well as ions typically observed in SE-SNe, such as \ion{Sc}{2}, \ion{Ti}{2}, and \ion{Si}{2} \citep[e.g.][]{SN2020cpg_Medler2021MNRAS.506.1832M, SN2020acat_Medler2022MNRAS.513.5540M, Zou-2026-ApJ}. The best-fit synthesized spectra of SN~2017gkk, along with the observed spectra, are shown in Figure~\ref{fig:specid}. The models fit our data well, and the prominent P~Cygni profiles they produce are consistent with the scenario of expanding ejecta above a photosphere.

Our first spectrum ($+$0.26~days) matches well with a synthesized spectrum with photospheric velocity $\sim12800\,\mathrm{km\,s^{-1}}$, and photospheric temperature  $\sim5\,\mathrm{kK}$. In the blue region blended metal lines due to \ion{Ca}{2} H\&K, \ion{Fe}{2}, \ion{Ti}{2} are reproduced reasonably well. A broad H$\alpha$ feature around 6200~\AA\ with a developing \ion{Si}{2} line in the blue wing, weak \ion{He}{2} and \ion{Ca}{2} NIR features are also well reproduced.

Due to the lower luminosity of the SN, the SNR in the later data ($+$28.7~days and $+$58.9~days from $V$-band maximum) is lower than that of the peak spectrum. Consequently, we can only identify some of the prominent features. The best-fit photospheric velocity decreases to $\sim9700\,\mathrm{km\,s^{-1}}$, following the velocity of \ion{Fe}{2} and indicating the photosphere recession. The broad absorption feature at $\sim$\,6200 \AA\ gradually becomes narrower and shallower. Furthermore, the components of this P~Cygni feature gradually become distinct. As indicated by the \textsc{SYNAPPS} results, the blue wing contribution from \ion{Si}{2} becomes isolated, suggesting that the ejecta are becoming transparent. The evolution of the velocity and profile of the $\sim$\,6200~\AA\ absorption blue wing differs from that of H$\alpha$. This suggests that they originate from different layers and are unlikely to be a high-velocity H$\alpha$ component, consistent with the findings of \cite{SN2022ngb_Zhao2025A&A}. As the SN evolves, the \ion{He}{1} lines become dominant, displaying clear P~Cygni profiles in the spectra at $+$28.7~days and $+$58.9~days from $V$-band maximum. This behavior is consistent with the spectroscopic evolution of SNe~IIb \citep{Filippenko1988F}. Metal lines at shorter wavelengths are still present; however, identifying individual lines is difficult due to the low SNR. Only the prominent \ion{Fe}{2}~$\lambda$5169~\AA\ and \ion{Fe}{2}~$\lambda$5018~\AA\ lines are identifiable. As the ejecta become optically thin, the emission of the \ion{Ca}{2} NIR triplet gradually becomes prominent, suggesting that the event is in transition to the nebular phase.

\begin{figure*}
\centering
\includegraphics[width=\textwidth]{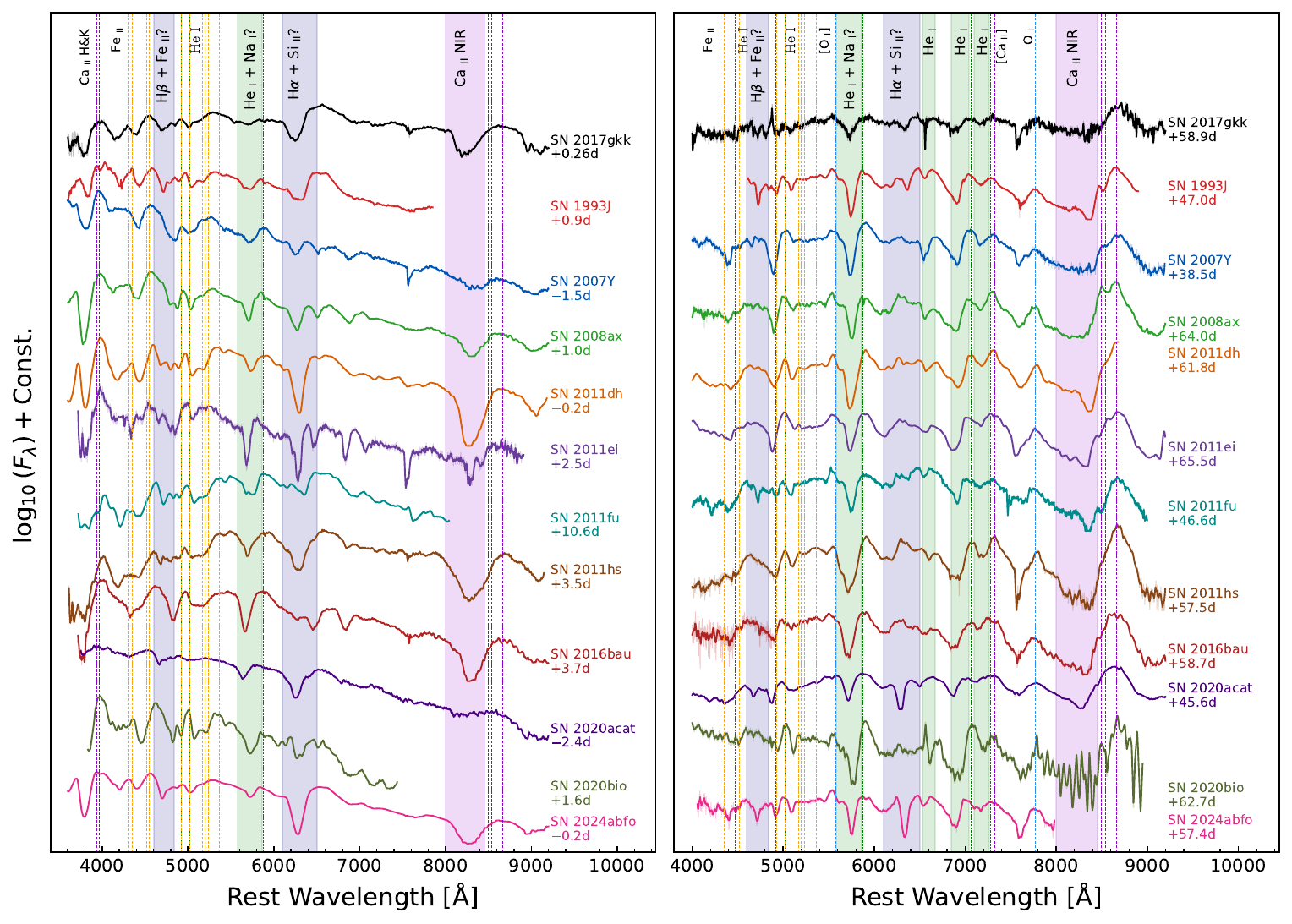}
\caption{Spectral comparison of SN~2017gkk with other Type IIb/Ib events: SNe~1993J, 2007Y, 2008ax, 2011ei, 2011fu, 2011hs, 2016bau, 2020acat, 2020bio and SN~2024abfo. The left and right panels show the near-maximum and the post-maximum spectra in the transition phase, respectively. Phases below each event are also indicated.}
\label{fig:speccmp}
\end{figure*}

\subsection{Comparisons to other SNe IIb/Ib}\label{sec:spec:cmp}

\begin{figure}[htbp]
\centering
\includegraphics[width=\columnwidth]{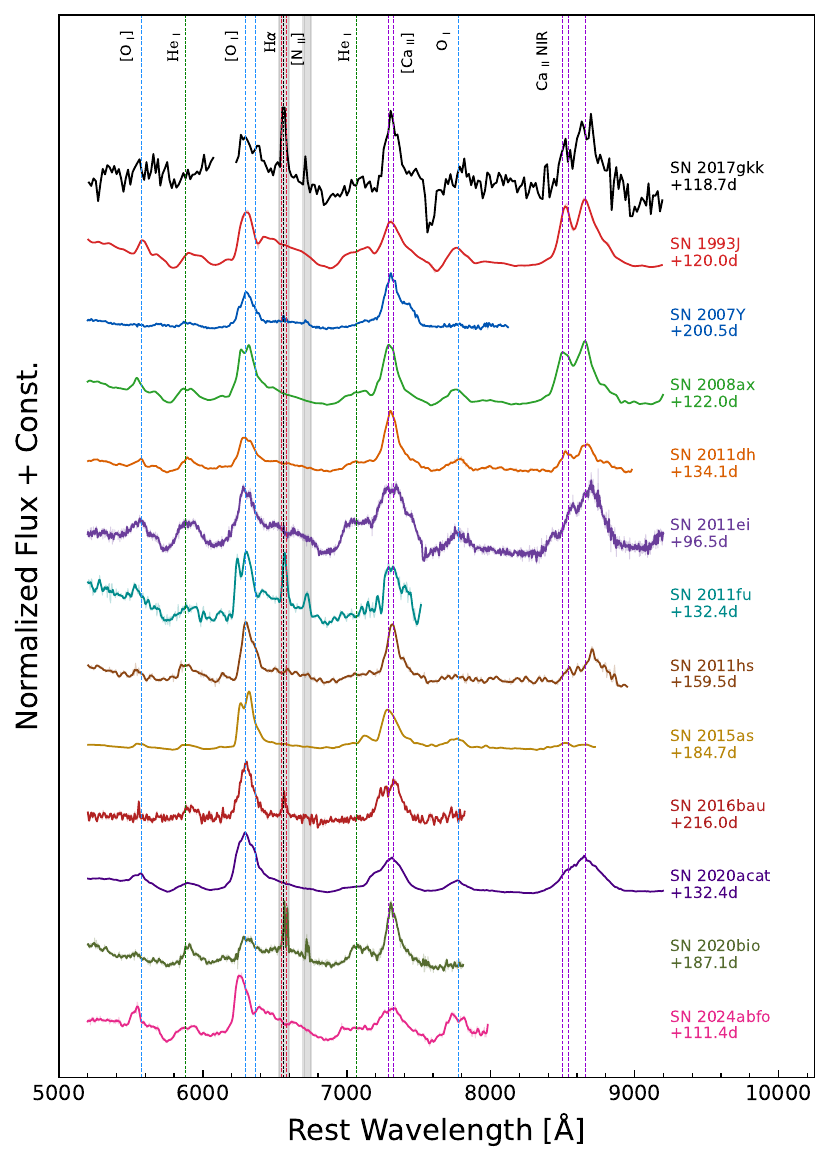}
\caption{Comparisons of the spectrum of SN~2017gkk with other Type IIb and Ib events as mentioned previously. The black dashed line represents the H$\alpha$ line, while red lines represent the [\ion{N}{2}] lines. The gray regions represent the narrow emission lines from \ion{H}{2} regions of host galaxy (H$\alpha$, [\ion{N}{2}] and [\ion{S}{2}]).}
\label{fig:speccmpnebul}
\end{figure}

We compared the spectra of SN~2017gkk with those of other well-studied IIb and a few low-luminosity Ib events from the literature. Given the absence of early-phase spectra for SN~2017gkk, we compared the spectra at peak brightness (+0.26 days relative to $V$-band maximum), the transition-phase spectrum (+58.9 days after $V$-band maximum), and the nebular-phase spectrum (+118.7~days after $V$-band maximum). SN~2017gkk generally follows the spectral evolution typical of SNe~IIb. The evolution of hydrogen and helium lines indicates a progenitor with a largely stripped envelope and helium-rich ejecta. These comparisons are presented in Figure~\ref{fig:speccmp} and Figure~\ref{fig:speccmpnebul}.

The spectral comparison at +0.26 days relative to the $V$-band maximum reveals features typical of Type II-like events. In the case of SN~2017gkk, the P~Cygni profiles are prominent. The spectrum is dominated by hydrogen lines, alongside prominent \ion{Ca}{2} NIR and H\&K lines. The metal lines in the blue region exhibit behavior generally consistent with the comparison sample. The P~Cygni profiles of \ion{He}{1} begin to emerge. We observe that the P~Cygni absorption around $\sim6200$~\AA\ for SN~2017gkk, primarily attributed to H$\alpha$, indicates a slightly higher velocity than that of SN~1993J, SN~2011dh \citep{SN2011dh_Sahu2013MNRAS.433....2S}, and SN~2011fu \citep{SN2011fu_Kumar2013MNRAS.431..308K}. This velocity is comparable to that of SN~2008ax and SN~2020acat, suggesting a relatively high ejecta velocity. This characteristic is also observed in the H$\beta$ line and \ion{He}{1} lines. Furthermore, the absorption profile around $\sim$\,6200~\AA\ exhibits a distinct, relatively broad structure that differs from those seen in SN~1993J and SN~2011fu. This feature can be explained by the fact that the low ejecta mass implies a highly mixed core enriched with $^{56}$Ni and Fe \citep{Taddia2018A&A...609A.136T}, whereas mixing within the outer H and He envelopes remains inefficient \citep{Jerkstrand2015A&A...573A..12J}. These mechanisms indicate distinct and clear H$\alpha$ and H$\beta$ lines, with separate and prominent lines from core ejecta such as \ion{Fe}{2} and \ion{Si}{2}, which is consistent with our observation.

The spectral comparison during the transition phase ($+$58.9~days relative to the $V$-band maximum) demonstrates general consistency with the comparison sample. In all cases, we observe that the H$\alpha$ and H$\beta$ lines fade, while the \ion{He}{1}~$\lambda5876$~\AA\ line becomes dominant, exhibiting a deep and clear absorption feature. The \ion{Si}{2} lines also become prominent, a characteristic common in He-rich events \citep[see, e.g.][]{Dessart2016MNRAS.458.1618D, Holmbo2023A&A...675A..83H}. Furthermore, as the ejecta become optically thin, forbidden emission lines such as [\ion{Ca}{2}]~$\lambda\lambda7291,7324$~\AA\ and [\ion{O}{1}]~$\lambda5577$~\AA\ begin to emerge. According to \cite{SN2015as_Gangopadhyay2018MNRAS}, the presence of [\ion{O}{1}]~$\lambda5577$~\AA\ indicates a high-temperature environment, suggesting that SN~2017gkk has not yet fully entered the nebular phase. Emission lines of \ion{He}{1} are prominent in the red region of the spectrum, alongside \ion{O}{1}~$\lambda7774$~\AA\ and the \ion{Ca}{2} NIR triplet. Collectively, these features confirm that SN~2017gkk is evolving from the photospheric phase to the nebular phase, and are consistent with most of the Type IIb events.

The spectral comparison during the nebular phase is presented in Figure~\ref{fig:speccmpnebul}. As the ejecta become optically thin, exposing the inner layers, nebular spectra provide valuable constraints on the properties of the progenitor and the explosion. We compared the spectrum of SN~2017gkk at $+$118.7~days ($\sim132$~days after explosion) with spectra of comparison events at similar epochs. Due to the low luminosity of the event, the nebular-phase spectrum of SN~2017gkk exhibits a lower SNR; nevertheless, prominent features remain identifiable. Additionally, SNe~IIb display significant diversity in their nebular-phase spectra. 

The nebular-phase spectra of all comparison events, including SN~2017gkk, are dominated by forbidden lines such as [\ion{O}{1}]~$\lambda\lambda6300,6364$~\AA\ and [\ion{Ca}{2}]~$\lambda\lambda7291,7324$~\AA. In addition, lines such as [\ion{O}{1}]~$\lambda5577$~\AA\ and \ion{O}{1}~$\lambda7774$~\AA, which are primarily associated with ionized oxygen recombination, are present in the spectra. The \ion{Ca}{2} NIR triplet is also observed as pure emission, as the absorption component has completely disappeared. The red wing of the \ion{Ca}{2} feature exhibits a distinct structure, likely due to the contribution of [\ion{C}{1}]~$\lambda8730$ \citep{SN2017iro_Kumar2022ApJ...927...61K, vanBaal2024MNRAS.532.4106V}, which is a common characteristic in late-time spectroscopy of SE-SNe. P~Cygni profiles become less prominent, while \ion{He}{1} emission indicates optically thin and He-rich ejecta. [\ion{N}{2}] lines are also detectable in all comparisons, which can be used to estimate the mass of the progenitor \citep{Jerkstrand2015A&A...573A..12J, Barmentloo2024MNRAS.533.1251B}. In the case of SN~2017gkk, the [\ion{O}{1}]~$\lambda5577$~\AA\ line is less prominent compared to that in the comparison events, likely obscured by spectral noise. This suggests that the ejecta became relatively cooler, and the ionized oxygen had almost disappeared. Furthermore, the \ion{He}{1} lines exhibit weak P~Cygni features. These characteristics indicate that SN~2017gkk entered the nebular phase with relatively low opacity \citep[see][]{SN2015as_Gangopadhyay2018MNRAS, SN2020acat_Medler2022MNRAS.513.5540M, SN2017iro_Kumar2022ApJ...927...61K}. This finding is consistent with our light-curve analysis, which indicates a low $M_{\mathrm{ej}}$ and a faster evolution compared to the other objects.

Since oxygen is synthesized by hydrostatic burning within the progenitor, the profile of [\ion{O}{1}]~$\lambda\lambda6300,6364$~\AA\ reflects the spatial distribution of oxygen \citep{Valenti2011MNRAS, SN2017iro_Kumar2022ApJ...927...61K, Fang2024NatAs}. Conversely, calcium is primarily formed through explosive burning via the chain $^{28}\mathrm{Si}(\alpha, \gamma){}^{32}\mathrm{S}(\alpha, \gamma){}^{36}\text{Ar}(\alpha, \gamma){}^{40}\mathrm{Ca}$, making it a major indicator of the explosion geometry \citep{vanBaal2024MNRAS.532.4106V}. We observe that the [\ion{O}{1}]~$\lambda\lambda6300,6364$~\AA\ emission of SN~2017gkk exhibits a blue-shifted component in the 6300~\AA\ peak and a split-peak profile, resembling the features seen in SN~2008ax, SN~2011ei, SN~2011fu, and SN~2015as. Regarding the [\ion{Ca}{2}]~$\lambda\lambda7291,7324$~\AA\ lines, SN~2017gkk displays a single, sharp peak, similar to SN~2011dh, SN~2011hs, and SN~2015as. 

We also identify [\ion{N}{2}] emission lines situated on the red wing of the [\ion{O}{1}]~$\lambda\lambda6300,6364$~\AA\ feature. Because the $^{14}\mathrm{N}(p,\gamma)^{15}\mathrm{O}$ reaction represents the bottleneck of the CNO cycle, $^{14}\mathrm{N}$ accumulates in the core during the main-sequence phase. During the subsequent He-burning phase, this $^{14}\mathrm{N}$ is effectively processed into $^{18}\mathrm{F}$ and subsequently decays to $^{18}\mathrm{O}$ \citep{Barmentloo2024MNRAS.533.1251B, SN2022ngb_Zhao2025A&A}, with more massive stars exhibiting a faster consumption rate. In the case of SN~2017gkk, the ejecta is helium-rich while lacking prominent hydrogen signatures, yet it exhibits distinct [\ion{N}{2}] emission. Furthermore, the presence of the [\ion{C}{1}]~$\lambda8730$ line on the red wing of the \ion{Ca}{2} NIR triplet implies that the progenitor retained a significant abundance of carbon. Both the nitrogen and carbon signatures indicate inefficient nuclear burning, which is characteristic of a lower-mass scenario. Collectively, these features point to a low-mass progenitor, a conclusion consistent with our previous analysis and the light curve modeling results. The detailed discussion and modeling of the nebular phase spectra of SN~2017gkk is presented in Sect.~\ref{sec:prog:nebspec_mdl}.

\begin{figure}[htbp]
\centering
\includegraphics[width=\columnwidth]{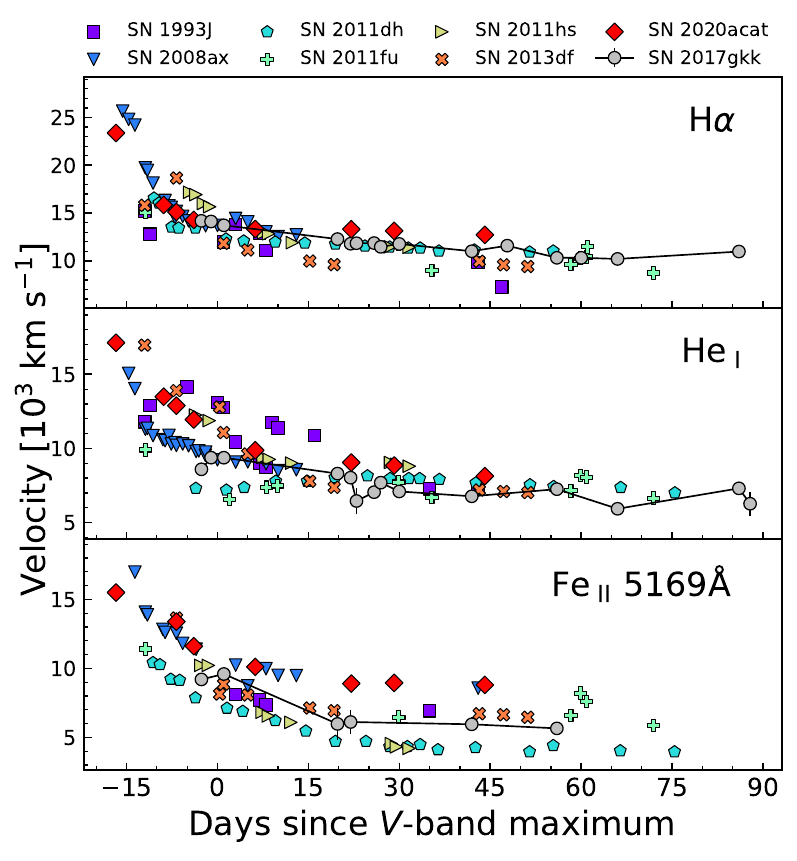}
\caption{Line velocities evolution of SN~2017gkk along with other comparisons. We measured and compared H$\alpha$, \ion{He}{1}~$\lambda5876$~\AA\ and \ion{Fe}{2}~$\lambda5169$~\AA. The comparisons data were either from WISeREP \citep{2012yaron} or literature \citep[e.g.][]{SN2011fu_Kumar2013MNRAS.431..308K, SN2011hs_Bufano2014MNRAS.439.1807B, SN2020acat_Medler2022MNRAS.513.5540M, SN2022ngb_Zhao2025A&A}.}
\label{fig:specvelocity}
\end{figure}

\subsection{line velocities}
\label{sec:spec:velocityevo}

We measured the spectral line velocities of SN~2017gkk and the comparison objects. The results are presented in Figure~\ref{fig:specvelocity}. We derived the velocities by performing a double Gaussian fit to the entire P~Cygni profile or a single Gaussian fit to the absorption component. An MCMC routine was employed to estimate robust uncertainties. We present the evolution of H$\alpha$, H$\beta$, \ion{He}{1}~$\lambda5876$~\AA, and \ion{Fe}{2}~$\lambda5169$~\AA. Since the S/N ratio in the blue region of the spectra is low, the results for H$\beta$ and \ion{Fe}{2} in the late phase are less reliable.

The line velocities of SN~2017gkk follow the general trend of the comparison objects. Although we lack early-phase spectroscopy, from the available data, the velocity evolution appears relatively flat and stable, and shows a generally slow decline. The velocity of H$\alpha$ declined from $14200\pm360\,\mathrm{km\,s^{-1}}$ at $+$0.26~days relative to the $V$-band maximum ($+$13.35 days from explosion) to $11000\pm650\,\mathrm{km\,s^{-1}}$ at $+$88.92~days relative to the $V$-band maximum ($+$102.01~days from explosion). This velocity traces the expansion of the H-rich envelope. SN~2017gkk exhibits typical velocities similar to the other objects in the sample. The velocity of H$\beta$ declined from $12000\pm480\,\mathrm{km\,s^{-1}}$ at $+$0.26~days to $4400\pm350\,\mathrm{km\,s^{-1}}$ at $+$29.84~days relative to the $V$-band maximum. The velocity of \ion{He}{1}~$\lambda5876$~\AA\ traces the surface of the He-rich core. For SN~2017gkk, the velocity measured $9400\pm500\,\mathrm{km\,s^{-1}}$ at $+$1.85~days relative to the $V$-band maximum, and subsequently declined to $6300\pm880\,\mathrm{km\,s^{-1}}$ at $+$90.76~days. This evolution follows the trend of SN~2008ax and SN~2011dh, but represents a slightly slower expansion than that of SN~1993J and SN~2020acat. 

Previous studies \citep[e.g.,][]{Schmutz-1990S, Branch-2002B, 2005bdessart} suggest that the photosphere is located between the metal-rich layer and the low-opacity hydrogen and helium layers. Consequently, transitions such as \ion{Fe}{2}, \ion{Ti}{2}, or \ion{Si}{2} serve as reliable probes of the photospheric velocity. Here, we considered the \ion{Fe}{2}~$\lambda5169$~\AA\ line as the tracer. The velocity of the \ion{Fe}{2}~$\lambda5169$~\AA\ line at $+$0.26 days was measured as $9200.9_{-351.1}^{+351.3}\,\mathrm{km\,s^{-1}}$; we derived this value precisely from the clear, high S/N spectrum obtained with NOT/ALFOSC. We adopted this value as the photospheric velocity at maximum light to break the degeneracy between $v_\mathrm{ej}$ and $M_{\mathrm{ej}}$ in the light curve fitting (see Sect.~\ref{sec:pho:lcmdl}).

\section{Discussions}\label{discussion}

\subsection{Light curve modeling}\label{sec:pho:lcmdl}
Here, we present our modeling of the bolometric light curve of SN~2017gkk. Unlike other Type IIb SNe such as SN~1993J and SN~2011fu, the bolometric light curve of SN~2017gkk does not exhibit the initial peak that is typically powered by the shock breakout and subsequent cooling phase. Therefore, its evolution can be effectively described by a single-component model,  
powered by the radioactive decay chain $^{56}$Ni$\rightarrow^{56}$Co$\rightarrow^{56}$Fe. We adopt the analytical framework of \cite{1982arnett, 1989arnett}, which assumes that in the early evolutionary phase, the explosion is spherically symmetric, the ejecta undergoes homologous expansion, and the $\gamma$-ray photons are fully trapped. Within this framework, the characteristic diffusion timescale ($\tau_{\mathrm{m}}$) and the kinetic energy ($E_{\mathrm{k}}$) of the ejecta are given by:
\begin{equation}
\tau_{\mathrm{m}} = \sqrt{\frac{2 \kappa_{\mathrm{opt}} M_{\mathrm{ej}}}{\beta c v_{\mathrm{ph}}}}, \;\;\; E_{\mathrm{k}} = \frac{3}{10} M_{\mathrm{ej}} v_{\mathrm{ph}}^2 \, .
\end{equation}
Here, $\beta \approx 13.8$ is a dimensionless integration constant in the Arnett-like model \citep{1989arnett, 2012Chatzopoulos}. The model parameters, specifically the ejecta mass ($M_{\mathrm{ej}}$) and kinetic energy ($E_{\mathrm{k}}$), are degenerate. To break this degeneracy, we incorporate an independent velocity measurement. We measure velocity  from the \ion{Fe}{2} \ $\lambda$5169~\AA\ line,  in the publicly available earliest spectrum, obtained $\sim$ 0.2 days after the $V$-band maximum \citep{Onori2017TNSCR.964....1O}. This velocity can be treated as representative of the photospheric velocity ($v_{\mathrm{ph}}$). We adopt a photospheric velocity of $v_{\mathrm{ph}}=9200 \pm350\,\mathrm{km\,s^{-1}}$ and approximate the diffusion timescale, $\tau_{\mathrm{m}}$, with the light curve rise time, which ranges from 13 to 17 days. For the optical opacity, we adopt $\kappa_{\mathrm{opt}}$ ranging from $0.07\,\mathrm{cm^{2}\,g^{-1}}$ to $0.195\,\mathrm{cm^{2}\,g^{-1}}$, a typical value for the inner ejecta of Type IIb supernovae \citep{Taddia2018A&A...609A.136T, Nagy2018ApJ, SN2020cpg_Medler2021MNRAS.506.1832M, SN2024abfo_Reguitti2025A&A...698A.129R}. With these parameters, we derive an ejecta mass in the range of $M_{\mathrm{ej}} = 0.61 - 2.95\,M_{\odot}$ and a corresponding kinetic energy of $E_{\mathrm{k}} = (0.31 - 1.49)\times10^{51}\,\mathrm{erg}$. The constraints derived from Arnett's rule are relatively broad; therefore, we treat these results as priors for our subsequent, more robust parameter estimation.

\begin{figure}
\centering
\includegraphics[width=\columnwidth]{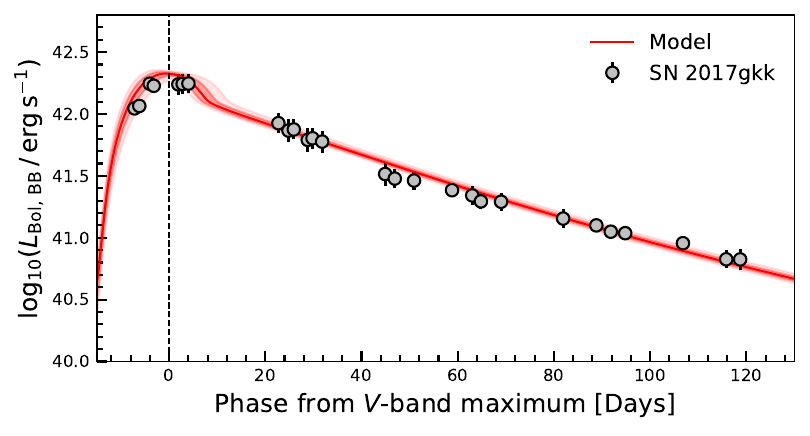}
\caption{Bolometric light curve fitting results based on a slightly modified Arnett-like model \citep{2012Chatzopoulos, Nagy2014}, obtained through MCMC sampling (see Section~\ref{sec:pho:lcmdl}). The solid lines indicate the best-fit models, and the lighter shaded regions correspond to the 1$\sigma$ and 2$\sigma$ confidence intervals.}
\label{fig:lcfitting_bollc}
\end{figure}

Estimating the $^{56}$Ni mass from the peak luminosity is unreliable due to the recombination effect (see Sect.~\ref{sec:pho:abslc_color} and Sect.~\ref{sec:pho:bollc}), which can provide an additional contribution to the luminosity at maximum light \citep{Nagy2014}. Hence, we constrain the $^{56}$Ni mass using the late-time tail of the light curve. As noted in Sect.~\ref{sec:pho:abslc_color}, the late-time decline is significantly faster than other comparisons like SN~1993J and SN~2011fu. This fast decline is attributed to the increasing transparency of the ejecta, which allows for significant $\gamma$-ray leakage. Adopting the approach of \citet{Clocchiatti1997ApJ} and \citet{2012Chatzopoulos}, the bolometric luminosity at these late phases can be modeled as the radioactive energy deposition rate, which is the total energy input attenuated by the escape of $\gamma$-rays:
\begin{equation}
\begin{split}
L_{\mathrm{Bol}}(t)&\approx L_{\mathrm{inp}}(t)\times\left(1-e^{-\tau_{\gamma}}\right)\\
&=M_{\mathrm{Ni}}\left[\left(\epsilon_{\mathrm{Ni}} - \epsilon_{\mathrm{Co}}\right)e^{-\frac{t}{\tau_{\mathrm{Ni}}}} + \epsilon_{\mathrm{Co}}e^{-\frac{t}{\tau_{\mathrm{Co}}}}\right]\\
&\quad \times\left(1-e^{-\tau_{\gamma}}\right)\;.
\end{split}
\end{equation}
In this model, $\tau_\gamma$ is the $\gamma$-ray optical depth. By assuming a constant average $\gamma$-ray opacity ($\kappa_\gamma$) and homologous free expansion (with density profile $\rho(x, t)=\rho_{00}\eta(x)R_0^{3}R(t)^{-3}$), the optical depth simplifies to a function of $t^{-2}$:
\begin{equation}
\begin{split}
\tau_{\mathrm{\gamma}}&=\kappa_{\mathrm{\gamma}}\rho_{00}R_0^{3}R(t)^{-2}\int_{0}^{1}\mathrm{d}x'\,\eta(x') \\
&\approx \frac{\kappa_{\mathrm{\gamma}}M_{\mathrm{ej}}I_{\mathrm{M}}^{\mathrm{0^{th}}}}{4\pi I_{\mathrm{M}}^{\mathrm{2^{nd}}}v_{\mathrm{ej}}^{2}}\times t^{-2} = A_\mathrm{\gamma} \times t^{-2}\;.
\end{split}
\end{equation}

The constants $I_{\mathrm{M}}^{0^{\mathrm{th}}}$ and $I_{\mathrm{M}}^{2^{\mathrm{nd}}}$ represent the zeroth and second moments of the dimensionless density profile. Because the estimate of $M_\mathrm{ej}$ is uncertain, we adopt a more direct approach for the preliminary estimate by treating the comprehensive term $A_\mathrm{\gamma}$, which encapsulates the ejecta's physical properties, as a single free parameter. Fitting our model to the bolometric light curve for $t > 30$ days yields a $^{56}$Ni mass of $M_{\mathrm{Ni}}=0.068\pm0.006\,M_\odot$. We note that $\gamma$-ray leakage is intrinsically linked to the ejecta density profile, and we discuss this in the following detailed fitting, along with constraints on other parameters.

We model the inner, helium-rich component of the bolometric light curve using the modified Arnett-like model \citep[see,][]{2012Chatzopoulos, NagyVinko2016}, as a lack of early-time data prevents constraining the outer hydrogen-rich shell. We adopt the typical parameters that have been used in other well-studied SNe~IIb \citep[e.g.,][]{2014morales, SN2015as_Gangopadhyay2018MNRAS, NagyVinko2016}. Thus, the initial radius and thermal energy are fixed to $R_0 = 2\times10^{11}\,\mathrm{cm}$ and $E_{\mathrm{th,\,init}} = 0.1\times10^{51}\,\mathrm{erg}$. We apply a Gaussian prior on the photospheric velocity, $v_{\mathrm{ej}}\sim \mathcal{N}(9200,\,351)\,\mathrm{km\,s^{-1}}$, to break the $M_{\mathrm{ej}}$--$v_{\mathrm{ej}}$ degeneracy. The optical opacity is fixed at $\kappa_{\mathrm{opt}}=0.07\,\mathrm{cm^2\,g^{-1}}$ \citep{Taddia2018A&A...609A.136T, SN2024abfo_Reguitti2025A&A...698A.129R}. The best-fit model is shown in Figure~\ref{fig:lcfitting_bollc}, and the posterior result is presented in Table~\ref{apptab:lcfitting}.

The best-fit model yields a rise time of $\sim$\,15 days, which is consistent with our previous estimation and comparable to SN~2020acat \citep[16.5~days;][]{SN2020acat_Medler2022MNRAS.513.5540M, SN2020acat_Ergon2024A&A...683A.241E} and SN~2011hs \citep[16.5 days;][]{SN2011hs_Bufano2014MNRAS.439.1807B}. The posterior ejecta mass for SN~2017gkk is $M_{\mathrm{ej}}=1.38_{-0.31}^{+0.48}\,M_{\odot}$. This agrees with low-mass cases typical of fast-evolving events like SN~2011ei \citep[$\sim$\,1.6\,$M_\odot$;][]{SN2011ei_Milisavljevic2013ApJ...767...71M} and SN~2011hs \citep[$\sim$\,1.8\,$M_\odot$;][]{SN2011hs_Bufano2014MNRAS.439.1807B}. The corresponding kinetic energy can be derived by $E_{\mathrm{k}}= (0.5/I_{\mathrm{M}}^{\mathrm{2^{nd}}})M_{\mathrm{ej}}v_{\mathrm{ej}}^{2}\times I_{\mathrm{M}}^{4^{\mathrm{th}}}$, considering our $M_\mathrm{ej}$ and exponential density profile \citep{SN2011fu_Kumar2013MNRAS.431..308K}, the best-fit kinetic energy is $0.69_{-0.17}^{+0.25}\times10^{51}\,\mathrm{erg}$, which is consistent with a less energetic event but with a relatively high energy-to-mass ratio.

The $^{56}$Ni mass derived for SN~2017gkk is $0.076_{-0.024}^{+0.032}\,M_\odot$, which is consistent with our previous estimate based on late-time $\gamma$-ray leakage. This value is consistent with those moderate luminous events like SN~2013df and SN~2015as \citep{2014morales, SN2015as_Gangopadhyay2018MNRAS}. These events also exhibit similar peak bolometric luminosities. The $\gamma$-ray leakage in SN~2017gkk is also significant, with a timescale of $59.57_{-2.02}^{+2.10}\,\mathrm{days}$. This timescale is considerably shorter than that of SN~2017ckj \citep[$\sim$80\, days;][]{SN2017ckj_Li2025A&A251022989L} and is consistent with the rapid decline rate of the light curve for SN~2017gkk. Since $\kappa_{\mathrm{\gamma}}$ depends on $M_\mathrm{ej}$, which in turn is affected by our choice of the optical opacity $\kappa_{\mathrm{opt}}$, we can only constrain $\kappa_{\mathrm{\gamma}}$ to a wide range. For $M_{\mathrm{ej}}=1.38_{-0.31}^{+0.48}\,M_{\odot}$, the corresponding $\kappa_{\mathrm{\gamma}}=0.033_{-0.010}^{+0.011}\,\mathrm{cm^{2}\,g^{-1}}$.

\subsection{Nature of the progenitor}\label{sec:prog}

In this section, we constrain and discuss the progenitor properties of SN~2017gkk based on the spectral modeling of the nebular phase and the discovery of the progenitor from the archival HST images by \citet{Niu2024ApJ...970L...9N}.  

\subsubsection{Nebular spectra modeling}\label{sec:prog:nebspec_mdl}

The latest nebular phase spectrum of SN~2017gkk was obtained at $\sim$\,118 days after $V-$band maximum, corresponding to $\sim$\,130 days from the estimated explosion epoch. We note that the [\ion{O}{1}]~$\lambda5577$ line is faint, suggesting that the ejecta was cooling. To verify whether the ejecta was optically thin, we fitted the [\ion{O}{1}]~$\lambda\lambda6300,6364$ feature using a two-component Gaussian model with a baseline. The fitting results yield an [\ion{O}{1}]~$\lambda6300$ to [\ion{O}{1}]~$\lambda6364$ flux ratio of $2.84_{-0.42}^{+0.49}$. This value is consistent with the theoretical prediction of $\sim3$ \citep{1988spyromilio}, suggesting that the ejecta are indeed nearly optically thin. We observed that the [\ion{O}{1}]~$\lambda\lambda6300,6364$ profiles are not perfectly Gaussian, as we briefly discussed in Sect.~\ref{sec:spec:cmp}; instead, they exhibit an asymmetry feature.

The nebular spectra represent the inner part of the ejecta and have a lot to do with the progenitor (see Sect.~\ref{sec:spec:cmp}). Since oxygen is mainly synthesized during the hydrostatic burning, the mass of oxygen can be a rough probe of the progenitor mass. \cite{1986uomoto} provides a relation to calculate the neutral oxygen mass, which is suitable for the high electron density and low temperature region:
\begin{equation}
M_{\mathrm{O}}=10^8d_\mathrm{L}^{2}f([\mathrm{O_{\,I}}])\times\mathrm{exp}\left(\frac{2.28}{T_4}\right)\,M_{\odot}\,.
\end{equation}

Here, $d_\mathrm{L}$ is the distance to the SN in Mpc, for SN~2017gkk, $d_\mathrm{L}=19.048\pm0.593\,\mathrm{Mpc}$. The $T_4$ is the dimensionless temperature of the oxygen emission region and is represented as $T_4 = T\times10^{-4}\,\mathrm{K^{-1}}$. Since the [\ion{O}{1}]~$\lambda5577$ becomes insignificant (which is lower than the noise level) and \ion{O}{1}~$\lambda7774$~\AA\, which is caused by recombination, tentatively becomes faint, we assume that all the oxygen is neutral and the environment agrees with the low temperature and high electron density. We measure the flux of [\ion{O}{1}]~$\lambda\lambda6300,6364$~\AA\ by applying a multi-component Gaussian fitting. The fitting result shows $f([\mathrm{O_{\,I}}])=(2.228_{-0.163}^{+0.183})\times10^{-15} \,\mathrm{erg\,s^{-1}\,cm^{-2}}$. Taking $T_4=0.4$, we result the $M_{\mathrm{O}}\approx0.023\,M_\odot$. This result shows a low oxygen mass, which corresponds to a low-mass progenitor. However, \cite{SN2020acat_Medler2022MNRAS.513.5540M} suggests that since the $f([\mathrm{O_{\,I}}])$ is evolving with time, it is not suitable to use the flux as a probe of the oxygen and progenitor mass. 

Since the calcium was mainly synthesized in the explosion and the oxygen was mainly synthesized in the hydrostatic burning phase, the ratio between [\ion{Ca}{2}] and [\ion{O}{1}] can be a better probe of the progenitor mass \citep{1989fransson}. From the integration of the spectra, the flux ratio of [\ion{Ca}{2}] and [\ion{O}{1}] is $1.686_{-0.130}^{+0.145}$. This ratio is much higher than SN~1993J ($\sim0.5$) and SN~2008ax ($\sim0.9$; \citealt{2009taubenberger}), suggesting that the progenitor of SN~2017gkk should be a low-mass helium star with a thin hydrogen envelope, with $M_{\mathrm{preSN}}\lesssim3\,M_\odot$. These results show that the progenitor of SN~2017gkk was more likely stripped in a binary system rather than a single WR-like progenitor stripped by strong stellar winds.

\begin{figure}[htbp]
\centering
\includegraphics[width=\columnwidth]{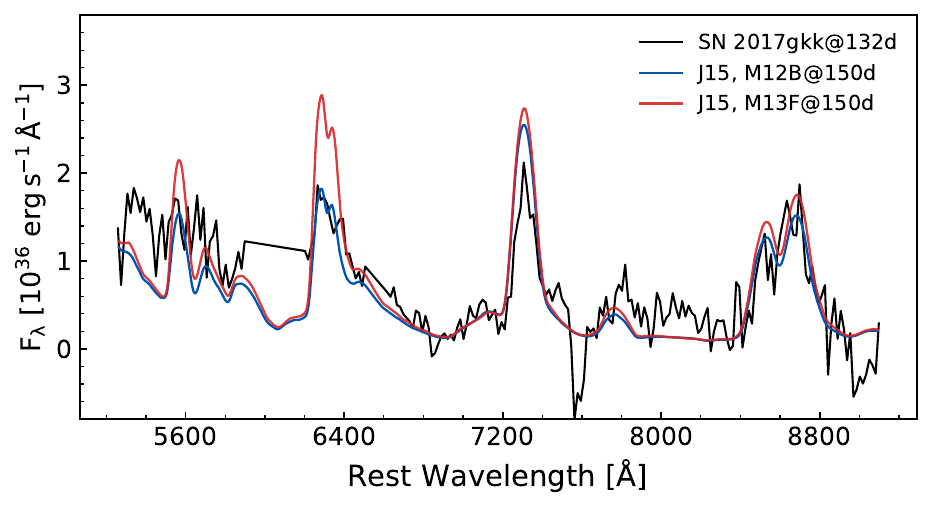}
\caption{Nebular phase spectra of SN~2017gkk over-plotted with synthetic model generated by \cite{Jerkstrand2015A&A...573A..12J} (\texttt{J15}, hereafter). Models with different physical parameters are shown with different colors. All the phases are calculated from the estimated explosion epoch.}
\label{fig:specmodel}
\end{figure}

To place better constraints on the progenitor system and the explosion scenario, we compared our nebular phase spectrum with the NLTE radiative-transfer models computed by \cite{Jerkstrand2015A&A...573A..12J}. To facilitate a direct comparison, we converted both the observed and model spectra to flux density and normalized them by the instantaneous radioactive power, accounting for $\gamma$-ray leakage and epoch differences. Additionally, artifacts around 6100~\AA\, were excised to strictly constrain the spectral matching.

\begin{table}[htbp]
    \centering
    \small
    \caption{Model parameters for selected models \citep[12B and 13F,][]{Jerkstrand2015A&A...573A..12J}.}
    \begin{tabularx}{\columnwidth}{ccccc}
        \hline
        Model & $M_{\rm ZAMS}$ & Mixing & $\text{e}^+$ & Molecule \\
              & ($M_\odot$)    &        &              &           \\ 
        \hline
        12B & 12 & Strong & Non-local & No \\
        13F & 13 & Strong & Non-local & No \\
        \hline
    \end{tabularx}
    \label{tab:specmdl}
\end{table}

Various model spectra are plotted in Figure~\ref{fig:specmodel}. Details of the models can be found in Table~\ref{tab:specmdl}. It is inferred that the best-fitting model is \texttt{M12B}, corresponding to a progenitor with $M_{\mathrm{ZAMS}}\approx12\,M_\odot$. The corresponding $M_{\mathrm{ej}}$ for these models is 1.7~$M_\odot$, which aligns well with our independent light curve modeling. Collectively, these comparisons point toward a low-mass progenitor. Notably, the emission on the red wing of the [\ion{O}{1}] feature is attributed to [\ion{N}{2}], \ion{Fe}{2}, and \ion{Fe}{1} \citep{Jerkstrand2015A&A...573A..12J, Barmentloo2024MNRAS.533.1251B, SN2022ngb_Zhao2025A&A}. This spectral feature is prominent in SN~2017gkk, further suggesting a low core mass, as massive progenitors would likely consume more $^{14}$N when the He starts to burn (see Sect.~\ref{sec:spec:cmp}).

\subsubsection{Progenitor identification with HST}

The pre-explosion progenitor of SN~2017gkk was observed with HST, and the archival images were investigated by \cite{Niu2024ApJ...970L...9N}. The progenitor was identified as a YSG with an initial mass of $\sim16\,M_\odot$. However, our light curve and spectral analysis indicate a lower pre-supernova mass of $M_\mathrm{preSN}\sim3\,M_\odot$, corresponding to a zero-age main sequence mass of $M_\mathrm{ZAMS}\sim12-13\,M_\odot$. This discrepancy might arise from the specific evolutionary path of the progenitor of SN~2017gkk. In the low-mass scenario, the progenitor is likely to have been stripped via binary interaction. Furthermore, a progenitor with $M_\mathrm{preSN}\sim3\,M_\odot$ could evolve from a wide range of $M_\mathrm{ZAMS}$. 

Additionally, \cite{Niu2024ApJ...970L...9N} suggested that SN~2017gkk was surrounded by a dense and dusty CSM, implying a high extinction of $A_\mathrm{V}^{\mathrm{CSM}}\sim3.05\,\mathrm{mag}$. This research also reported narrow emission lines in a spectrum obtained $+0.26$~days relative to the $V$-band maximum (see Section~\ref{sec:spec}). However, we note that SN~2017gkk is located deep within the host galaxy, where narrow emission from underlying \ion{H}{2} regions is not negligible. 
We measured the narrow emission lines in our spectrum taken at $+22.7$~days from the $V$-band maximum. By applying a Gaussian profile fit, we derived\footnote{Since spectral smoothing has little effect on the early-time spectra, it does not significantly degrade the effective spectral resolution. In particular, the H$\alpha$ and [\ion{N}{2}] lines remain clearly resolved in the spectroscopic evolution plot (Figure~\ref{fig:specevo}).} an FWHM of $\sim1100\,\mathrm{km\,s^{-1}}$, which is similar to the CSM velocity reported by \cite{Niu2024ApJ...970L...9N}, suggesting that the narrow lines in the spectra of SN~2017gkk are more likely to originate from the host rather than the CSM interaction. This result implies that the narrow lines observed at $+0.26$~days might not originate from CSM interaction.
Moreover, a study of SN~2017ckj \citep{SN2017ckj_Li2025A&A251022989L} demonstrates that CSM interaction features typically appear only within the first few days after the explosion. For SN~2017gkk, the phase of $+0.26$~days relative to maximum is likely too late for such interaction features to persist. Finally, since any dusty CSM would be rapidly destroyed following the explosion, the large extinction associated with the CSM should decrease over time. Consequently, applying such a large, constant extinction to the light curve modeling would be problematic.

However, \cite{Baer-Way2024ApJ...964..172B} suggested the late-time CSM interaction based on HST photometry extending to $\sim$\,1485 days after the explosion, as the light curve decline rate of SN~2017gkk is slower than that expected from the radioactive decay of $^{56}\mathrm{Co}\rightarrow^{56}\mathrm{Fe}$. This indicates that the progenitor system of SN~2017gkk underwent a period of mass loss of $\sim500-1000$~years before the explosion, which can be primarily attributed to mass transfer in a binary system \citep{Maeda2015ApJ, SN2024abfo_deWet2025arXiv250711131D}. Given the lack of very late-phase spectral data, we are limited to commenting on the late-phase CSM interaction. In such observations, we would expect to detect box-like H$\alpha$ emission lines similar to those observed in SN~1993J and SN~2013df.

\section{Summary}\label{sec:summary}

In this work, we investigated the Bessell $BVRI$ broad-band photometric and low-resolution spectroscopic data of Type IIb SN~2017gkk. Our observations (supplemented with archival data) cover both the maximum light and nebular phases. To obtain the true SN magnitudes, image subtraction was applied, as the SN was deeply embedded within the host.
Our analysis infers that SN~2017gkk is a moderately luminous and fast-evolving Type IIb SN, resembling SN~2011hs, SN~2011ei, and SN~2024aecx. By applying the ANN method, we estimated the explosion epoch and derived a $V$-band rise time of $13.09^{+0.71}_{-0.74}$ days (Section~\ref{sec:pho}). The $V$-band light curve also exhibits a rapid decline, with a decay rate of $2.35\pm0.06\,\mathrm{mag}\,(100\,\mathrm{days})^{-1}$ after 25 days from $V-$band maximum luminosity. We modeled the bolometric light curve utilizing the modified Arnett model, incorporating recombination and $\gamma-$ray leakage effects (see Section~\ref{sec:pho:lcmdl}). This analysis yields a low ejecta mass of $M_{\mathrm{ej}}=1.38_{-0.31}^{+0.48}\,M_{\odot}$ and a synthesized $^{56}$Ni mass of $0.076_{-0.024}^{+0.032}\,M_\odot$. The explosion energy is relatively low, estimated at $E_\mathrm{KE}=0.69_{-0.17}^{+0.25}\times10^{51}\,\mathrm{erg}$. Furthermore, $\gamma-$ray leakage is prominent in SN~2017gkk, characterized by a leakage parameter of $A_\gamma=3548_{-237}^{+254}\,\mathrm{day^2}$.

The low-resolution spectral data of SN~2017gkk presented in this work provide a coverage between $\sim$+2 to $\sim$+119 days relative to the $V$-band maximum. SN~2017gkk follows the typical spectroscopic evolutionary path of Type IIb SNe, with Balmer lines appearing in the early phases and gradually fading. These lines disappear by $\sim90$~days, suggesting that SN~2017gkk transitions into a Type Ib-like object, which is a normal characteristic of Type IIb events. 
Utilizing \textsc{SYNAPPS} for line identification, we observed that metal lines (e.g., \ion{Fe}{2} and \ion{Ti}{2}) are prominent in the blue region of the spectra, contributing significantly to the opacity during the transitional phase. Additionally, \ion{He}{1} lines emerge as the photosphere retreats into the ejecta.

In the nebular phase spectrum, the [\ion{O}{1}]~$\lambda\lambda6300,6364$ lines become prominent and exhibit a split-peak feature, indicating an asymmetric explosion or a toroidal distribution of oxygen. The flux ratio between [\ion{Ca}{2}] and [\ion{O}{1}]~$\lambda\lambda6300,6364$ suggests that the progenitor of SN~2017gkk was a low-mass star. Through comparison with NLTE spectral models (see Section~\ref{sec:prog}), we constrain the progenitor mass to $M_{\mathrm{preSN}}\sim3\,M_\odot$, corresponding to a zero-age main sequence mass of $M_{\mathrm{ZAMS}}\sim12-13\,M_\odot$, indicating the SN~2017gkk progenitor was likely in a binary system. 

In our environmental analysis and direct observations of SN~2017gkk, we find no strong evidence of early CSM interaction. However, late-time HST data reveal a slower decline rate, suggesting that CSM interaction became significant around $\sim1500$~days after the explosion. This implies a history of mass loss driven by binary mass transfer $\sim1000$~years prior to the explosion. To verify this scenario, detailed spectroscopic investigation of nebular phase data and sophisticated spectral modeling are required. However, due to a lack of data, we are limited to the present findings.

\section{Software and third party data repository citations} \label{sec:soft}

\software{astropy \citep{astropy:2022}, pandas \citep{reback2020pandas}, numpy \citep{harris2020array}, scipy \citep{2020SciPy-NMeth}, Jupyter-notebook \citep{Kluyver2016jupyter}.}

\section*{Acknowledgments}
BK acknowledges support from the ``Special Project for High-End Foreign Experts", Xingdian Funding from Yunnan Province, the National Key Research and Development Program of China (2024YFA1611603) and the ``Key Laboratory of Survey Science of Yunnan Province” (project No. 202449CE340002). G.C.A. acknowledges support from the Indian National Science Academy (INSA) under their Senior Scientist Programme. M.S. acknowledges financial support provided under the National Post Doctoral Fellowship (N-PDF; File Number: PDF/2023/002244) by the Science \& Engineering Research Board (SERB), Anusandhan National Research Foundation (ANRF), Government of India. 

We are grateful to the observers at HCT who provided their valuable time to monitor this event, and also thank the staff of IAO, Hanle, and CREST, Hosakote, that made these observations possible. The Weizmann interactive supernova data repository (WISeREP) - \url{http://wiserep.weizmann.ac.il} is also acknowledged. This research has made use of the NASA/IPAC Extragalactic Database (NED), which is operated by the Jet Propulsion Laboratory, California Institute of Technology, under contract with the National Aeronautics and Space Administration. This research made use of \textsc{RedPipe}\footnote{\url{https://github.com/sPaMFouR/RedPipe}} \citep{2021redpipe}, an assemblage of data reduction and analysis scripts written by AS. We thank the anonymous referee for his/her insightful comments and suggestions that helped significantly improve the manuscript.

\appendix

\section{Complementary tables \& figures}\label{app:extratabfig}

\begin{table*}[htbp]
\centering
\caption{Calibrated magnitudes of secondary standard stars in the field of SN~2017gkk. Star IDs are shown in Figure~\ref{fig:sn-field}.}
\label{apptab:calibphotometryinfo}
\begin{tabular}{cccccc}\hline
Star& $B$              & $V$              & $R$              & $I$              \\
ID  & (mag)            & (mag)            & (mag)            & (mag)            \\ \hline
1   & 17.94 $\pm$ 0.02 & 17.26 $\pm$ 0.01 & 16.83 $\pm$ 0.01 & 16.49 $\pm$ 0.02 \\ 
2   & 16.64 $\pm$ 0.01 & 15.83 $\pm$ 0.01 & 15.35 $\pm$ 0.01 & 14.97 $\pm$ 0.01 \\ 
3   & 18.66 $\pm$ 0.03 & 17.50 $\pm$ 0.02 & 16.74 $\pm$ 0.01 & 16.12 $\pm$ 0.01 \\ 
4   & 15.90 $\pm$ 0.01 & 15.23 $\pm$ 0.01 & 14.83 $\pm$ 0.01 & 14.48 $\pm$ 0.01 \\ 
5   & 16.31 $\pm$ 0.01 & 15.55 $\pm$ 0.01 & 15.09 $\pm$ 0.01 & 14.68 $\pm$ 0.01 \\ 
6   & 17.22 $\pm$ 0.01 & 16.39 $\pm$ 0.01 & 15.90 $\pm$ 0.01 & 15.50 $\pm$ 0.01 \\ 
7   & 16.61 $\pm$ 0.01 & 16.10 $\pm$ 0.01 & 15.76 $\pm$ 0.01 & 15.42 $\pm$ 0.01 \\ 
8   & 17.08 $\pm$ 0.01 & 15.72 $\pm$ 0.01 & 14.87 $\pm$ 0.01 & 14.12 $\pm$ 0.01 \\ 
9   & 18.02 $\pm$ 0.02 & 17.16 $\pm$ 0.01 & 16.66 $\pm$ 0.01 & 16.20 $\pm$ 0.01 \\ 
10  & 17.36 $\pm$ 0.01 & 16.17 $\pm$ 0.01 & 15.48 $\pm$ 0.01 & 14.93 $\pm$ 0.01 \\ 
11  & 16.66 $\pm$ 0.01 & 15.82 $\pm$ 0.01 & 15.33 $\pm$ 0.01 & 14.90 $\pm$ 0.01 \\ \hline
\end{tabular}
\end{table*}
\vspace{-2em}
\begin{table}[htbp]
    \centering
    \caption{Apparent light curve parameters of SN~2017gkk}
    \begin{tabular}{cccccc}
        \hline
        Filter & System & $t_{peak}$ & $t_{rise}^\dagger$ & $m_{peak}$ & $\gamma_{25<t<80}$ \\
        && (MJD) & (days) & (mag)  & ($\mathrm{mag\times(100~days)^{-1}}$)\\
        \hline
        $B$&Vega& $ 57997.30 \pm 1.25 $& $ 12.41^{+1.32}_{-1.34} $ & $ 16.70 \pm 0.04 $ & $0.73\pm0.10$ \\
        $V$&Vega& $ 57997.98 \pm 0.56 $& $ 13.09^{+0.71}_{-0.74} $ & $ 15.82 \pm 0.02 $ & $2.35\pm0.06$ \\
        $R$&Vega& $ 57997.62 \pm 0.56 $& $ 12.73^{+0.71}_{-0.74} $ & $ 15.26 \pm 0.02 $ & $2.78\pm0.11$ \\
        $I$&Vega& $ 57996.89 \pm 0.62 $& $ 12.00^{+0.76}_{-0.78} $ & $ 14.92 \pm 0.02 $ & $2.62\pm0.07$ \\
        \hline
    \end{tabular}\\
    \label{apptab:apparentlc}
    \vspace{1em}
    \scriptsize 
    \raggedright
    $^\dagger$: The rising time is calculated from the maximum epoch of each band to the explosion epoch. Considering the poor coverage in early data and the explosion epoch is estimated by extrapolation, which could be highly uncertain. Thus, the rising time here can only be a reference.
\end{table}
\begin{table}
    \centering
    \caption{Bolometric light curve fitting results of SN~2017gkk}
    \begin{tabular}{cccc}\hline
        Parameters & Unit & Prior & Posterior \\ \hline
        $f_{\mathrm{Ni}}\,^{\ddagger}$ & -- & $\mathcal{U}(-3,\,1)$ & $0.055_{-0.012}^{+0.013}$ \\
        $M_{\mathrm{ej}}\,^{\ddagger}$ & $M_\odot$ & $\mathcal{U}(-1,\,0.5)$ & $1.38_{-0.31}^{+0.48}$ \\
        $v_{\mathrm{ej}}$ & $\mathrm{km\,s^{-1}}$ & $\mathcal{N}(9200,\,351)$ & $9120_{-410}^{+430}$ \\
        $A_{\mathrm{\gamma}}\,^{\ddagger}$ & $\mathrm{day^2}$ & $\mathcal{U}(3,\,5)$ & $3548_{-237}^{+254}$ \\
        $T_{\mathrm{exp}}$ & $\mathrm{day}$ & $\mathcal{U}(-15,\,-1)$ & $-7.15_{-1.33}^{+0.99}$ \\
        $\sigma_\mathrm{add}\,^{*\,\ddagger}$ & $\mathrm{dex}$ & $\mathcal{U}(-5,\,1)$ & $0.025_{-0.004}^{+0.006}$ \\
        \hline
    \end{tabular}
    \label{apptab:lcfitting}
    \vspace{1em}
    \scriptsize 
    \raggedright

    $^{*}$: The uncertainty of the bolometric light curve is not well estimated, as the fitting can't take the uncertainty as the weight. Thus, we add a jitter, which represents the systematic uncertainty.\\
    $^{\ddagger}$: The priors of these parameters are uniform in logarithmic space with base 10.
\end{table}

\begin{figure}
\centering
\includegraphics[width=\columnwidth]{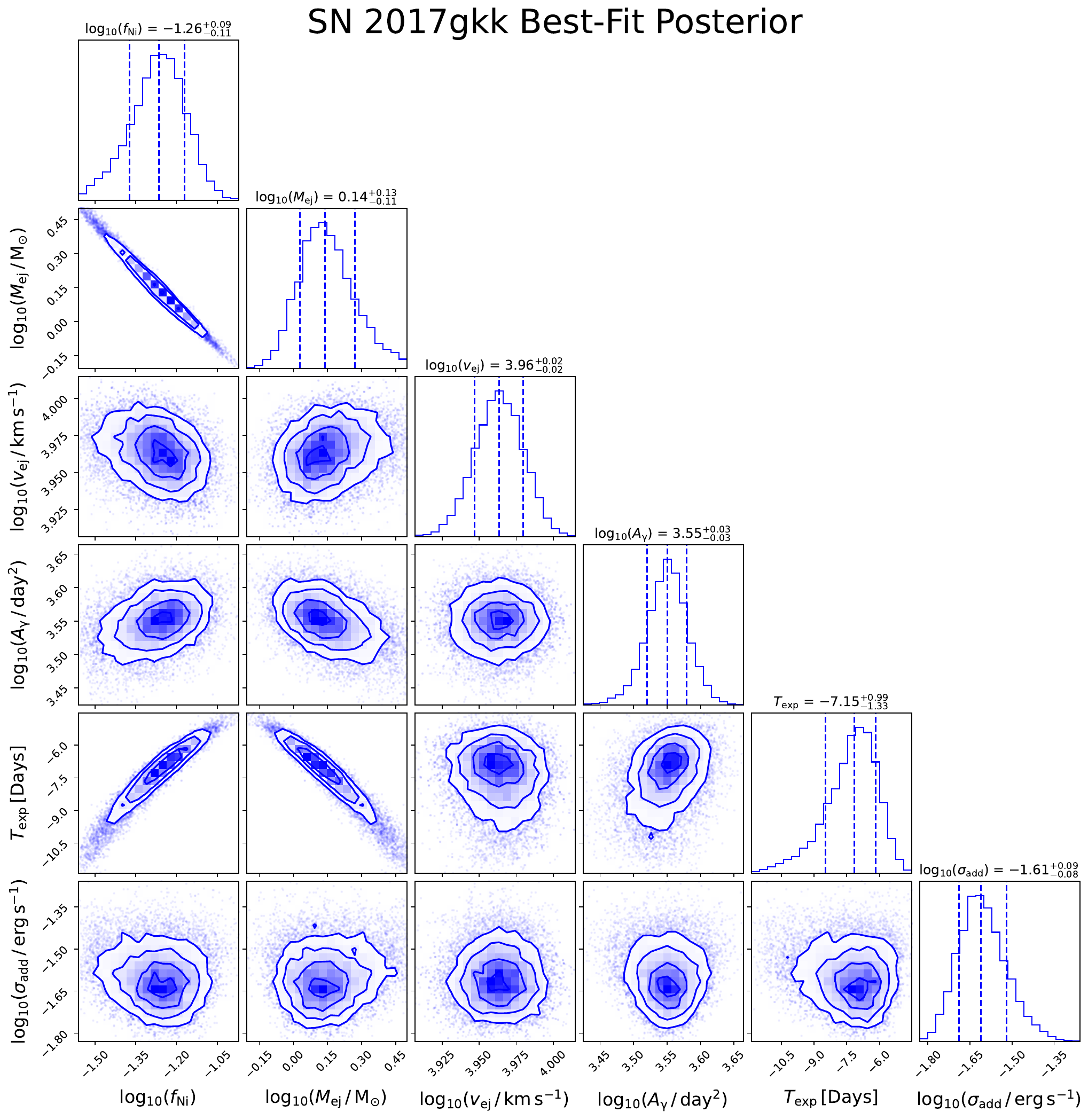}
\caption{Posteriors distribution of bolometric light curve fitting of SN~2017gkk.}\label{appfig:corner}
\end{figure}

\clearpage
\section{Observation Logs}
\begin{longtable}[htbp]{@{\extracolsep{0.1em}}cccccccc}
\caption{Log of spectroscopic observations obtained from HCT.}
\small
\label{apptab:spectroscopy_log} \\
\hline \hline
Date & MJD & Phase$^a$ & Telescope & Grism & Spectral range & Resolution & Exp. time \\ 
  &   & (days) &   &        & (\AA)    &           & (s)           \\ 
\hline
2017-09-03 & 57999.83 & +1.85 & HCT & Gr.7/Gr.8 & 3500–7800, 5200–9250 & R$\sim$1000; 1200 & 1200 \\
2017-09-05 & 58001.94 & +3.96 & HCT & Gr.7/Gr.8 & 3500–7800, 5200–9250 & R$\sim$1000; 1200 & 1500 \\
2017-09-24 & 58020.68 & +22.7 & HCT & Gr.7/Gr.8 & 3500–7800, 5200–9250 & R$\sim$1000; 1200 & 2400 \\
2017-09-26 & 58022.85 & +24.87 & HCT & Gr.7/Gr.8 & 3500–7800, 5200–9250 & R$\sim$1000; 1200 & 2700 \\
2017-09-27 & 58023.82 & +25.84 & HCT & Gr.7/Gr.8 & 3500–7800, 5200–9250 & R$\sim$1000; 1200 & 2700 \\
2017-09-30 & 58026.72 & +28.74 & HCT & Gr.7/Gr.8 & 3500–7800, 5200–9250 & R$\sim$1000; 1200 & 2700 \\
2017-10-01 & 58027.82 & +29.84 & HCT & Gr.7/Gr.8 & 3500–7800, 5200–9250 & R$\sim$1000; 1200 & 2700 \\
2017-10-04 & 58030.87 & +32.89 & HCT & Gr.7/Gr.8 & 3500–7800, 5200–9250 & R$\sim$1000; 1200 & 2400 \\
2017-10-16 & 58042.83 & +44.85 & HCT & Gr.7/Gr.8 & 3500–7800, 5200–9250 & R$\sim$1000; 1200 & 2700 \\
2017-10-22 & 58048.68 & +50.7 & HCT & Gr.7/Gr.8 & 3500–7800, 5200–9250 & R$\sim$1000; 1200 & 2700 \\
2017-10-30 & 58056.87 & +58.89 & HCT & Gr.7/Gr.8 & 3500–7800, 5200–9250 & R$\sim$1000; 1200 & 2100 \\
2017-11-03 & 58060.89 & +62.91 & HCT & Gr.7/Gr.8 & 3500–7800, 5200–9250 & R$\sim$1000; 1200 & 2700 \\
2017-11-09 & 58066.88 & +68.9 & HCT & Gr.7/Gr.8 & 3500–7800, 5200–9250 & R$\sim$1000; 1200 & 2700 \\
2017-11-29 & 58086.90 & +88.92 & HCT & Gr.7/Gr.8 & 3500–7800, 5200–9250 & R$\sim$1000; 1200 & 2700 \\
2017-12-01 & 58088.74 & +90.76 & HCT & Gr.7/Gr.8 & 3500–7800, 5200–9250 & R$\sim$1000; 1200 & 2700 \\
2017-12-02 & 58089.82 & +91.84 & HCT & Gr.7/Gr.8 & 3500–7800, 5200–9250 & R$\sim$1000; 1200 & 2700 \\
2017-12-29 & 58116.65 & +118.67 & HCT & Gr.8 & 5200–9250 & R$\sim$1200 & 2700 \\
\hline
\multicolumn{8}{l}{{$^a$Phases are relative to $V$-band maximum light. }} \\
\end{longtable}
\raggedbottom

\begin{longtable}[htbp]{@{\extracolsep{0.1em}}cccccccc}
\caption{Log of photometric observations.}
\label{apptab:photometrydata} \\
\hline \hline
Date        &  MJD      & Phase$^{\ast}$ & $B$          & $V$          & $R$                & $I$ & Telescope            \\
(yyyy-mm-dd)& &(d)             & (Vega mag)        & (Vega mag)        & (Vega mag)              & (Vega mag)          \\
\hline
2017-08-25 & 57991.3 & $-$6.7 & 17.26 $\pm$ 0.14 & 16.38 $\pm$ 0.11 & 15.77 $\pm$ 0.12 & 15.30 $\pm$ 0.11 & 60/90-Schmidt$^{\dagger}$\\
2017-08-26 & 57992.3 & $-$5.7 & 17.25 $\pm$ 0.14 & 16.29 $\pm$ 0.10 & 15.70 $\pm$ 0.12 & 15.21 $\pm$ 0.11 & 60/90-Schmidt$^{\dagger}$\\
2017-08-28 & 57994.5 & $-$3.5 & 16.73 $\pm$ 0.08 & 15.92 $\pm$ 0.05 & 15.34 $\pm$ 0.06 & 14.97 $\pm$ 0.06 & 60/90-Schmidt$^{\dagger}$\\
2017-08-29 & 57995.3 & $-$2.7 & 16.81 $\pm$ 0.09 & 15.91 $\pm$ 0.07 & 15.32 $\pm$ 0.07 & 14.94 $\pm$ 0.07 & 60/90-Schmidt$^{\dagger}$\\
2017-09-03  & 57999.9 & +1.9  & 16.78 $\pm$ 0.07 & 15.87 $\pm$ 0.04 & 15.35 $\pm$ 0.02 & 15.03 $\pm$ 0.03 & HCT \\
2017-09-04  & 58000.8 & +2.8  &         --       &          --      & 15.30 $\pm$ 0.04 &         --       & HCT\\
2017-09-05  & 58002.0 & +4.0  & 16.77 $\pm$ 0.17 & 15.86 $\pm$ 0.02 & 15.28 $\pm$ 0.03 & 14.96 $\pm$ 0.03 & HCT\\
2017-09-24  & 58020.7 & +22.7 & 19.10 $\pm$ 0.03 & 17.15 $\pm$ 0.02 & 16.25 $\pm$ 0.02 & 15.71 $\pm$ 0.04 & HCT\\
2017-09-26  & 58022.8 & +24.8 & 18.97 $\pm$ 0.02 & 17.23 $\pm$ 0.01 & 16.34 $\pm$ 0.02 & 15.78 $\pm$ 0.01 & HCT\\
2017-09-27  & 58023.9 & +25.9 & 19.15 $\pm$ 0.02 & 17.29 $\pm$ 0.01 & 16.37 $\pm$ 0.01 & 15.81 $\pm$ 0.03 & HCT\\
2017-09-30  & 58026.8 & +28.8 & 19.13 $\pm$ 0.15 & 17.31 $\pm$ 0.08 & 16.37 $\pm$ 0.08 & 15.87 $\pm$ 0.10 & HCT\\
2017-10-01  & 58027.8 & +29.8 & 19.13 $\pm$ 0.04 & 17.40 $\pm$ 0.01 & 16.51 $\pm$ 0.01 & 15.92 $\pm$ 0.02 & HCT\\
2017-10-03  & 58029.8 & +31.8 & 19.22 $\pm$ 0.05 & 17.46 $\pm$ 0.02 & 16.58 $\pm$ 0.01 & 15.99 $\pm$ 0.02 & HCT\\
2017-10-16  & 58042.9 & +44.9 & 19.31 $\pm$ 0.03 & 17.78 $\pm$ 0.01 & 16.97 $\pm$ 0.01 & 16.36 $\pm$ 0.02 & HCT\\
2017-10-18  & 58044.9 & +46.9 & 19.27 $\pm$ 0.02 & 17.82 $\pm$ 0.01 & 17.11 $\pm$ 0.02 & 16.42 $\pm$ 0.01 & HCT\\
2017-10-22  & 58048.9 & +50.9 & 19.30 $\pm$ 0.04 & 17.88 $\pm$ 0.01 & 17.12 $\pm$ 0.01 & 16.48 $\pm$ 0.01 & HCT\\
2017-10-30  & 58056.8 & +58.8 & 19.41 $\pm$ 0.03 & 18.07 $\pm$ 0.01 & 17.33 $\pm$ 0.01 & 16.66 $\pm$ 0.02 & HCT\\
2017-11-03  & 58061.0 & +63.0 &        --        & 18.11 $\pm$ 0.02 & 17.42 $\pm$ 0.02 & 16.81 $\pm$ 0.01 & HCT\\
2017-11-05  & 58062.8 & +64.8 &        --        & 18.21 $\pm$ 0.03 & 17.57 $\pm$ 0.02 & 16.92 $\pm$ 0.02 & HCT\\
2017-11-09  & 58067.0 & +69.0 &        --        &        --        & 17.51 $\pm$ 0.02 & 16.94 $\pm$ 0.02 & HCT\\
2017-11-22  & 58079.9 & +81.9 &        --        & 18.50 $\pm$ 0.02 & 17.94 $\pm$ 0.01 & 17.29 $\pm$ 0.03 & HCT\\
2017-11-29  & 58086.8 & +88.8 & 19.98 $\pm$ 0.04 & 18.73 $\pm$ 0.02 & 18.09 $\pm$ 0.02 & 17.35 $\pm$ 0.01 & HCT\\
2017-12-02  & 58089.8 & +91.8 & 19.85 $\pm$ 0.05 & 18.76 $\pm$ 0.03 & 18.26 $\pm$ 0.03 & 17.71 $\pm$ 0.05 & HCT\\
2017-12-05  & 58092.8 & +94.8 &        --        & 18.85 $\pm$ 0.03 & 18.20 $\pm$ 0.02 & 17.67 $\pm$ 0.02 & HCT\\
2017-12-17  & 58104.8 & +106.8&        --        & 19.03 $\pm$ 0.02 & 18.44 $\pm$ 0.01 & 17.78 $\pm$ 0.02 & HCT\\
2017-12-26  & 58113.8 & +115.8& 20.59 $\pm$ 0.05 & 19.34 $\pm$ 0.02 & 18.86 $\pm$ 0.02 & 18.02 $\pm$ 0.02 & HCT\\
2017-12-29  & 58116.7 & +118.7&        --        & 19.32 $\pm$ 0.03 & 18.86 $\pm$ 0.02 & 18.02 $\pm$ 0.03 & HCT\\
2018-01-11  & 58129.7 & +131.7&        --        &          --      & 19.10 $\pm$ 0.03 & 18.36 $\pm$ 0.09 & HCT\\
2018-01-21  & 58139.8 & +141.8& 21.26 $\pm$ 0.10 & 20.38 $\pm$ 0.07 & 19.16 $\pm$ 0.03 & 19.03 $\pm$ 0.18 & HCT\\
2018-02-13  & 58162.7 & +164.7&         --       &          --      & 19.53 $\pm$ 0.02 & 19.43 $\pm$ 0.08 & HCT\\
2018-04-07  & 58215.7 & +217.7&         --       &          --      & 20.04 $\pm$ 0.04 & 20.33 $\pm$ 0.09 & HCT\\
\hline
\hline
\multicolumn{8}{l}{{$^a$ Phases are relative to $V$-band maximum light.}}\\
\multicolumn{8}{l}{{$^{\dagger}$ These data points were obtained from \url{https://www.astronomerstelegram.org/?read=10686.}}} \\
\end{longtable}
\raggedbottom

\bibliography{2-SN2017gkk-n}
\bibliographystyle{aasjournalv7}

\end{document}